\documentclass[a4paper,twocolumn,10pt,accepted=2024-02-26]{quantumarticle}
\pdfoutput=1
\usepackage[english]{babel}
\usepackage[T1]{fontenc}
\usepackage[numbers, sort&compress]{natbib}

\usepackage{graphicx}
\usepackage{dcolumn}
\usepackage{bm}
\usepackage{physics}
\usepackage{amsmath, amssymb, amsfonts, amsthm, dsfont}
\usepackage{color}
\usepackage{url}
\usepackage[hidelinks]{hyperref}
\usepackage[capitalise]{cleveref}
\usepackage{microtype}
\usepackage[normalem]{ulem}

\newcommand{\ii}{\mathrm{i}}
\newcommand{\id}{\mathds{1}}

\newcommand{\overbar}[1]{\mkern 1.5mu\overline{\mkern-1.5mu#1\mkern-1.5mu}\mkern 1.5mu}

\begin{document}

\title{Custom Bell inequalities from formal sums of squares}

\author{Victor Barizien}
\affiliation{Université Paris Saclay, CEA, CNRS, Institut de physique théorique, 91191 Gif-sur-Yvette, France}
\author{Pavel Sekatski}
\affiliation{Département de Physique Appliquée, Université de Genève, 1211 Genève, Suisse}
\author{Jean-Daniel Bancal}
\affiliation{Université Paris Saclay, CEA, CNRS, Institut de physique théorique, 91191 Gif-sur-Yvette, France}

\begin{abstract}
Bell inequalities play a key role in certifying quantum properties for device-independent quantum information protocols. It is still a major challenge, however, to devise Bell inequalities tailored for an arbitrary given quantum state. 
Existing approaches based on sums of squares provide results in this direction, but they are restricted by the necessity of first choosing measurement settings suited to the state. Here, we show how the sum of square property can be enforced for an arbitrary target state by making an appropriate choice of nullifiers, which is made possible by leaving freedom in the choice of measurement. Using our method, we construct simple Bell inequalities for several families of quantum states, including partially entangled multipartite GHZ states and qutrit states. In most cases we are able to prove that the constructed Bell inequalities achieve self-testing of the target state. We also use the freedom in the choice of measurement to self-test partially entangled two-qubit states with a family of settings with two parameters. Finally, we show that some statistics can be self-tested with distinct Bell inequalities, hence obtaining new insight on the shape of the set of quantum correlations.
\end{abstract}

\maketitle

\section{Introduction}

\noindent One of the most striking characteristic of quantum theory is the fact that it does not admit a hidden variable description that is local, a property usually referred to as nonlocality~\cite{Brunner14}. This feature is at the heart of numerous phenomenons and applications ranging from quantum paradoxes~\cite{Brassard05} to device-independent information processing, which enables the certification of quantum properties such as entanglement~\cite{Werner89,Bancal11} and randomness~\cite{Colbeck06,Colbeck11,Pironio10} without relying on an underlying description of the apparatuses involved~\cite{Mayers04,Acin07}.

Bell inequalities are the tool of choice to study nonlocality: the violation of a Bell inequality certifies nonlocality, and any nonlocal behavior can be highlighted by a Bell inequality. Furthermore, quantum applications based on nonlocality are validated by Bell expressions: the length of a key distributed between two parties in a device-independent quantum key distribution protocol, for instance, is a direct function of a Bell score~\cite{Ekert91,Acin07,Nadlinger22,Zhang22,Liu22}. The nonlocal properties of quantum states and measurements can thus be investigated by studying Bell inequalities. However, apart from a few notable exceptions, Bell inequalities suited to quantum states and/or quantum measurements of interest are generally not known.

Indeed, Bell inequalities were initially studied for their ability to distinguish local from nonlocal behaviors. Given a number of measurement settings and possible outcomes, the set of local behaviors forms a polytope which is singled out by its facets -- tight Bell inequalities~\cite{Pitowsky89}. Since a polytope has a finite number of facets, tight Bell inequalities are naturally of particular interest. The best known one is the Clauser-Horne-Shimony-Holt (CHSH) inequality, which involves two binary measurement settings per party~\cite{Clauser69}. This inequality has been studied extensively, and it is known to be maximally violated by performing complementary measurements on a two-qubits maximally entangled state $\ket{\phi^+}=(\ket{00}+\ket{11})/\sqrt{2}$. For this reason, the CHSH inequality is well suited to the study of this state's nonlocality. However, the direct relation between tight Bell inequalities and states of general interest, like maximally entangled states, essentially ends here.

Notably, the local polytope emerging when considering two measurements with three possible outcomes, a natural scenario for measurements on a three-dimensional quantum system, also has a unique new facet given by the so-called CGLMP inequality~\cite{Kaszlikowski02,Collins02}. Remarkably, this inequality is maximally violated by the two-qutrits state $\ket{\psi_\text{CGLMP}}\propto 2\ket{00}+(\sqrt{11}-\sqrt{3})\ket{11}+2\ket{22}$, which is non-maximally entangled~\cite{Acin02,Navascues07,Ioannou22}. This leaves open the question of identifying a Bell inequality suited to the maximally entangled state of two qutrits, or more generally to any other quantum state of particular interest.

A number of works provided partial answers to this question by successfully constructing Bell inequalities that are maximally violated by particular target states. Examples include Bell inequalities maximally violated by the maximally entangled state of two qutrits~\cite{Ji08,Liang09,Lim10,Salavrakos17}, and by partially entangled two-qubit states~\cite{Acin12}. Unfortunately, these approaches rely heavily on the specific structure or symmetry of the target state and do not generalize easily to arbitrary situations.

Other works managed to obtain Bell inequalities suited to generic situations by focusing on specific applications of nonlocality. This includes inequalities bounding optimally the communication cost~\cite{Pironio03}, the min entropy~\cite{NietoSilleras14,Bancal14} or the von Neumann entropy~\cite{Brown23}. However, these approaches depend on the full probability distributions and rely on the Navascués-Pironio-Acín (NPA) hierarchy~\cite{Navascues07,Navascues08} to relate to quantum theory. They have thus mostly been considered numerically for given choices of measurements, requiring an a priori guess of the measurements that should be performed on the state of interest to obtain correlations on the boundary of the quantum set.

Further insight on the relation between Bell inequalities and quantum states came from an independent line of work which showed that the two can be more strongly connected than anticipated. Namely, it was found that some Bell inequalities are not only maximally violated by specific states, but that these states are sometimes the only ones able to achieve their maximal violation (together with fundamentally equivalent realizations, such as isometrically equivalent ones)~\cite{Popescu92,Mayers04}. When this is the case, it is said that the Bell inequality self-tests the quantum state and/or measurements. The self-testing property strongly supports the idea of associating to a quantum state the Bell inequalities that it can maximally violate.

Initially, the self-testing property was only observed in a few specific instances, including the CHSH inequality~\cite{Popescu92,McKague12}, but it has now been confirmed in an overwhelming number of cases~\cite{Supic20}. For instance, self-testing schemes have been found for all partially entangled states of two qubits through the tilted Bell inequality~\cite{Bamps15} as well as for maximally entangled states in arbitrary dimension~\cite{Sarkar21}. It is still unknown whether all pure entangled states can be self-tested, but substantial result have been obtained along these line, see~\cite{Coladangelo17,Wu14,Pal14,McKague14,Supic18,Baccari19,Supic22,Sarkar22}.

Note that some self-testing results are based on the knowledge of the full measurement statistics rather than solely on the maximal violation of a Bell inequality~\cite{Mayers04,Yang14,Coladangelo17,Supic18,Rai22}. Leaving aside the case of non-exposed points~\cite{Goh18,Chen23}, any result obtained in this way can also be certified by a Bell inequality due to the convex nature of the quantum set, but the appropriate Bell expression may be hard to find~\cite{Coladangelo18}. Whereas self-tests based on full statistics rely on numerous parameters, the ones based on a single quantity may be easier to use and lead to a wide range of applications, as in the case of partially entangled states self-tested from the tilted Bell inequality~\cite{Wu14,Sainz16,Coladangelo17,Supic18,Zwerger19}. Constructing simple Bell inequalities is thus relevant even for states already known to be self-testable.

Among the techniques developed for self-testing, sum of squares (SOS) play an important role~\cite{Bamps15,Supic16,Supic20}. Indeed, significant properties of a Bell expression such as its Tsirelson bound can be inferred from its sum of squares decomposition~\cite{Doherty08}. 

Sum of squares have also been used to construct Bell expressions from a fixed choice of quantum state and measurements~\cite{Salavrakos17,Augusiak19,Kaniewski19,Santos23}.
Finally, the self-testing property itself was used to construct Bell inequalities, potentially for arbitrary multipartite states~\cite{Zwerger19}.
While generic, this last methods also relies on the usage of a choice of the measurements. Therefore, no technique for constructing Bell expressions tailored to a specific quantum state is known that is really generic, analytical and flexible.

Here, we present a systematic method that enables the construction of Bell expressions for generic target quantum states with the guarantee that the state reaches the maximal value of the Bell expression. Our method is rooted in the idea that the state maximally violates the constructed Bell inequality. 
As an introduction to the construction of Bell expressions under this simple principle, we review the variational method, first introduced in~\cite{Pal14,Sekatski18}. To our knowledge, this method was not described exhaustively in the literature. 
We then discuss some of the limitations of the variational method before introducing our approach based on formal sums of squares (SOS). This approach is complementary to the variational method in the sense that it generally provides sufficient conditions for the maximal value of a Bell expression to be achieved by a target state, while the variational method only provides necessary conditions. Readers only interested in the formal SOS method may jump directly to \cref{sec:SOSmethod}. Finally, we apply our method to several cases, construct each time the corresponding Bell inequalities, and discuss the relation to self-testing.

\section{Conditions for the maximal violation of a Bell inequality}

\subsection{General definitions}\label{sec:generalDefs}

Consider a setting where $n$ parties share a global state $\ket{\psi}$ on which they may perform $m$ different local measurements with $k$ possible outcomes.
This experiment is described by the conditional probability distribution $\bm P = P(a_1,\dots, a_n|x_1,\dots, x_n)$ of observing the outcomes $a_i=1,\ldots,k$ given the possible measurement settings $x_i= 1,\dots,m$. A \emph{Bell expression} $\beta$ in this scenario is a linear map that associates a Bell score
\begin{equation}\label{eq:BellExpression}
\beta(\bm P) = \sum_{\bm a, \bm x}\alpha_{\bm a|\bm x} P(\bm a|\bm x)
\end{equation}
to every probability distribution $\bm P$~\cite{Rosset14}. Here, $\alpha_{\bm a|\bm x}$ are the Bell expression's coefficients and $\bm a=(a_1,\ldots,a_n)$, $\bm x=(x_1,\ldots,x_n)$ are vectors containing the outcomes and setting choices of all parties. 
The well-known CHSH Bell inequality $\beta\leq 2$ is a bound on the CHSH Bell expression $\beta$ defined by~\cite{Clauser69}
\begin{equation}\label{eq:CHSHBellExpression}
\alpha_{a_1,a_2|x_1,x_2}=(-1)^{a_1+a_2+(x_1-1)(x_2-1)}
\end{equation}
with $a_1,a_2,x_1,x_2=1,2$. We emphasize that a Bell expression \cref{eq:BellExpression} is an object acting on probability space and is fully defined by the Bell coefficients $\alpha_{\bm a|\bm x}$.

Given a choice of measurement projectors $\{\hat \Pi_{a|x}^{(i)}\}$ with $\hat \Pi_{a|x}^{(i)}\in\mathcal{L}(\mathcal{H}^{(i)})$ and $\sum_a \Pi_{a|x}^{(i)}=\id$, a Bell expression $\beta$ gives rise to a \emph{Bell operator}
\begin{equation}
\hat S = \sum_{\bm a,\bm x}\alpha_{\bm a|\bm x} \hat \Pi_{a_1|x_1}^{(1)}\otimes\ldots\otimes\hat\Pi_{a_n|x_n}^{(n)}
\end{equation}
which acts on the full Hilbert space $\mathcal{H}=\mathcal{H}^{(1)}\otimes\ldots\otimes\mathcal{H}^{(n)}$: $\hat S\in \mathcal{L}(\mathcal{H})$. In the case of binary outcomes ($k=2$), the measurements can also be described in terms of measurement operators $\hat M_x^{(i)}=\hat \Pi_{1|x}^{(i)}-\hat\Pi_{2|x}^{(i)}$ with eigenvalue $\pm 1$ for each party. The Bell operator can then be rewritten in the form
\begin{equation}\label{eq:BellOperator}
\hat S = \sum_{x_1,\dots,x_n \geq 0} c_{x_1,\dots,x_n}\, \hat M_{x_1}^{(1)}\otimes \dots \otimes \hat M_{x_n}^{(n)},
\end{equation}
where $c_{x_1,\dots,x_n}\in\mathbb{R}$ and we set $\hat M_0^{(i)}=\id$. Bell operators satisfy
\begin{equation}
\beta(\bm P)=\bra{\psi} \hat S \ket{\psi}
\end{equation}
for every state $\ket{\psi}$~\cite{Scarani01}. In the case of the CHSH expression, choosing the optimal Pauli measurements $\hat M^{(1)}_1=\hat Z_A$, $\hat M^{(1)}_2=\hat X_A$ and  $\hat M^{(2)}_{x_2}=(\hat Z_B - (-1)^{x_2} \hat X_B)/\sqrt{2}$) gives rise to the Bell operator
\begin{equation}\label{eq:XXZZOperator}
\hat S = (\hat X_A \hat X_B + \hat Z_A \hat Z_B)/2.
\end{equation}
Here we omit the tensor notation and refer to parties with the letters $A$ and $B$. This operator is a well known entanglement witness~\cite{Guehne09} and has the remarkable property of identifying the singlet state within the two-qubits state space as the only state with maximal eigenvalue.

Clearly, a Bell operator depends both on the Bell coefficients $\alpha_{\bm a|\bm x}$ and on the choice of measurements. Furthermore, it acts on the parties' Hilbert space (e.g.~on $\mathbb{C}^{2^n}$ when $\ket{\psi}$ is an $n$-qubit state) rather than on probabilities. Hence, for a fixed Bell expression $\beta$, different choices of measurements $\hat M_x^{(i)}$ give rise to different Bell operators $\hat S$. Similarly, a given Bell operator $\hat S$ gives rise to different Bell expressions $\beta$ depending on the chosen set of measurements $\hat M_x^{(i)}$. In the following, we explore the constraints that a maximal violation imposes on the relation between Bell expressions and Bell operators and use them to identify Bell expressions that are relevant to a target state $\ket{\psi}$.

\subsection{Variational method}

The strongest connection known to date between Bell expressions $\beta$ and quantum states $\ket{\psi}$ occurs when the maximal score of a Bell expression self-tests a specific state~\cite{Supic20}, i.e.~when its maximum quantum score is only compatible with $\ket{\psi}$ (up to redundant transformations). However, it is not clear how a Bell expression can be generically constructed to self-test an arbitrary state. For this reason, we now relax this problem and consider conditions on Bell expressions that are only necessary for self-testing $\ket{\psi}$. This is an easier task and it can be used to discard Bell expressions $\beta$ that have no chance of self-testing $\ket{\psi}$.

One such necessary condition is for the maximal value of $\beta$ to be achieved by measuring $\ket{\psi}$. 
In this case, a first rather naive observation is that there must be an implementation of the measurement operators $\hat M_x^{(i)}$ such that $\ket{\psi}$ is an eigenstate with maximal eigenvalue of the corresponding Bell operator \cref{eq:BellOperator}.

When $\ket{\psi}=\ket{\phi^+}$, an example of such an operator is $\hat S$ given in \cref{eq:XXZZOperator}: $\ket{\phi^+}$ is its only eigenstate with eigenvalue 1. Now, depending on the actual measurement operators $\hat M_x^{(i)}$, this Bell operator can correspond to various Bell expressions. Considering arbitrary qubit measurements in the $\hat X$-$\hat Z$ plane
\begin{subequations}\label{eq:XZmeas}
\begin{align}
\hat A_x &= \cos(a_x) \hat Z_A + \sin(a_x) \hat X_A\\
\hat B_y &= \cos(b_y) \hat Z_B + \sin(b_y) \hat X_B
\end{align}
\end{subequations}
with $a_x,b_y\in\mathbb{R}$, $\hat S$ can be rewritten in terms of the measurement operators directly as
\begin{equation}\label{eq:BellOpXXZZ}
\begin{split}
\hat S =& \Big[\cos(a_2-b_2) \hat A_1 \hat B_1 - \cos(a_2-b_1) \hat A_1 \hat B_2\\
& -  \cos(a_1-b_2) \hat A_2 \hat B_1 + \cos(a_1-b_1)\hat A_2 \hat B_2 \Big]\\
&\times\frac{1}{2\sin(a_1-a_2)\sin(b_1-b_2)},
\end{split}
\end{equation}
which allows one to write all such Bell expressions as
\begin{equation}\label{eq:BellExprXXZZ}
\begin{split}
\beta =& \Big[\cos(a_2-b_2) \langle A_1 B_1 \rangle - \cos(a_2-b_1)\langle A_1 B_2 \rangle\\
& -  \cos(a_1-b_2) \langle A_2 B_1 \rangle + \cos(a_1-b_1)\langle A_2 B_2 \rangle\Big]\\
&\times\frac{1}{2\sin(a_1-a_2)\sin(b_1-b_2)},
\end{split}
\end{equation}
where the notation $\langle A_x B_y\rangle$ without hats stands for
\begin{equation}\label{eq:CorrelatorsInProbabilities}
\langle A_x B_y\rangle=\sum_{a,b=1}^2 (-1)^{a+b}P(a,b|x,y).
\end{equation}

At this stage, we note that these expressions may not all be good candidates for Bell expressions maximized by $\ket{\psi}$ as implementations with other measurement settings may give rise to Bell operators with eigenvalues larger than $1$. Consider for example the Bell expression for the choices of parameter $a_1=0$, $a_2=\pi/2$, $b_y=-(-1)^y \pi/6$
\begin{equation} \label{eq:bellcandidatenotgood}
\beta =\frac{1}{2\sqrt{3}}\Big[\langle A_1 B_1 \rangle + \langle A_1 B_2 \rangle +  \sqrt{3} \langle A_2 B_1 \rangle - \sqrt{3} \langle A_2 B_2 \rangle\Big].
\end{equation}
By construction, its value for the maximally entangled state $\ket{\phi^+}$ with this choice of measurement is 1, but if Bob changes his measurements to $B_y = \cos(\pi/4) \Hat Z_B - (-1)^y \Hat X_B$, this value increases up to $\beta(\bm P) = \frac{\sqrt{2}}{2\sqrt{3}} (1+\sqrt{3})\simeq 1.12 > 1$. For these settings, measurements on a partially entangled state can also reach the value 1. We would thus like to refine the condition to avoid such cases.

So far, we only considered the relation between Bell operators and Bell expressions for a fixed choice of measurement settings, but if $\beta$ is maximized by $\ket{\psi}$, not only must $\ket{\psi}$ be a maximal eigenvector for the Bell operator corresponding to the chosen measurements, but the corresponding eigenvalue shall exceed the Bell score attainable for any other implementation. Ultimately, this should be the case even for implementations involving measurements of arbitrary dimension, which are arguably hard to parametrize. It should already be satisfied, however, in the qubit space. This case is easier to parametrize and constitutes a necessary condition. Even then, this condition is difficult to verify globally on the whole space as shown above. However, the condition takes a simple form when considering small perturbations of the ideal implementation.

To see this, consider the Bell expression \cref{eq:BellOpXXZZ} with measurements close to \cref{eq:XZmeas}, namely with
\begin{subequations}
\begin{align}
\hat A_x &\to \hat A_x + \delta_{A_x}(-\sin(a_x) \hat Z_A + \cos(a_x) \hat X_A)\\
\hat B_y &\to \hat B_y + \delta_{B_y}(-\sin(b_y) \hat Z_B + \cos(b_y) \hat X_B)
\end{align}
\end{subequations}
for small $\delta$s\footnote{In general, a perturbation of the ideal measurements operators can always be written as
\begin{equation}
\hat \Pi_{a|x}^{(i)} \mapsto e^{\ii h_x \delta} \hat \Pi_{a|x}^{(i)} e^{-\ii h_x \delta} \approx \hat \Pi_{a|x}^{(i)} + \delta \, \ii [h_x,  \hat \Pi_{a|x}^{(i)}]
\end{equation}
for $h_x = h_x^\dag$.
}. 
The derivatives with respect to $\delta_{A_x}$ and $\delta_{B_y}$ are given by 
\begin{small}
\begin{equation}
\begin{split}
    \frac{\partial \hat S}{\partial \delta_{A_1}} &=\frac{ (\cos (a_1)\hat{X}_A-\sin (a_1) \hat{Z}_A) (\cos(a_2) \hat{X}_B - \sin(a_2) \hat{Z}_B)}{2 \sin(a_1-a_2)} \\
     \frac{\partial \hat S}{\partial \delta_{A_2}} &=\frac{ (\cos (a_2)\hat{X}_A-\sin (a_2) \hat{Z}_A) (\sin(a_1) \hat{Z}_B-\cos(a_1) \hat{X}_B)}{2 \sin(a_1-a_2)} \\
     \frac{\partial \hat S}{\partial \delta_{B_1}} 
      &=\frac{  (\cos(b_2) \hat{X}_A - \sin(b_2) \hat{Z}_A)(\cos (b_1)\hat{X}_B-\sin (b_1) \hat{Z}_B)}{2 \sin(b_1-b_2)}\\
       \frac{\partial \hat S}{\partial \delta_{B_2}} &=\frac{  (\sin(b_1) \hat{Z}_A-\cos(b_1) \hat{X}_A)(\cos (b_2)\hat{X}_B-\sin (b_2) \hat{Z}_B)}{2 \sin(b_1-b_2)}.
\end{split}
\end{equation}\end{small}
The state $\ket{\phi^+}$ is a local optimum only if the expectation value of the resulting perturbed Bell operator remains unchanged to first order, i.e.~if
\begin{equation} \label{eq:firstorderphi+}
\begin{split}
\bra{\phi^+}\frac{\partial \hat S}{\partial \delta_{A_x}}\ket{\phi^+} &\propto \cos(a_1)\cos(a_2) + \sin(a_1)\sin(a_2) = 0\\
\bra{\phi^+}\frac{\partial \hat S}{\partial \delta_{B_y}}\ket{\phi^+} &\propto \cos(b_1)\cos(b_2) + \sin(b_1)\sin(b_2) = 0.
\end{split}
\end{equation}

In the range $-\frac{\pi}{2}\leq a_1,b_1 \leq\frac{\pi}{2}$, these conditions are equivalent to $a_2 = a_1 \pm \frac{\pi}{2}$ and $b_2 = b_1 \pm \frac{\pi}{2}$, i.e.~imposing that the measurements be complementary for Alice and Bob. Choosing $a_2=a_1+\pi/2$ and $b_2=b_1-\pi/2$, the corresponding Bell expression can be expressed as a function of a single parameter $c=b_1-a_1$:
\begin{equation}\label{eq:betac}
\begin{split}
\beta=&\cos(c) \langle A_1 B_1 \rangle+ \sin(c)\langle A_1 B_2 \rangle\\
&+ \sin(c)\langle A_2 B_1\rangle - \cos(c)\langle A_2 B_2\rangle.
\end{split}
\end{equation}
From this example, we see how  the condition of local optimality eliminates many Bell expressions candidates for $\ket{\phi^+}$. The resulting family contains CHSH as a special case, but also includes additional Bell expressions. One can verify that these expressions self-test the desired state for all $c\in(0,\pi/4]$. We comment later on reasons why specific values of $c$ might be more interesting than others in this particular example.

Note that \cref{eq:firstorderphi+} contains a redundancy: when the considered perturbation corresponds to a common rotation of both measurements $\delta_{A_1}=\delta_{A_2}$, it can be seen as a local unitary transformation of the state and the first order condition is always fulfilled. Thus, this condition is only sensitive to variations of the relative angle $\delta_A=\delta_{A_1}-\delta_{A_2}$ between the two measurements. The same holds on Bob's side, leaving one equality constraint per party. In general, parametrization up to a local unitary can be achieved by parametrizing for each party all measurements except one.

This example illustrates the usage of the variational principle to find a Bell expression maximized by a given quantum state. Here, we started from a specific Bell operator in \cref{eq:XXZZOperator} that is maximized by $\ket{\phi^+}$, but clearly, any other similar choice could have been made and the same procedure could be followed for an arbitrary state $\ket{\psi}$. We can thus formulate the variational method as follows:

\begin{enumerate}
    \item Choose a Bell operator $\Hat S$ which admits the target state $\ket{\psi}$ as its eigenstate with maximal eigenvalue.
    \item Parametrize the measurement bases, e.g.~$\hat M_{x}^{(i)}$ in the binary case, for each party.
    \item Define a corresponding parametrization of Bell expressions $\beta$ by expressing the Bell operator $\hat S$ in terms of the measurement operators $\hat M_{x}^{(i)}$.
    \item Consider a perturbation of the measurements
    \begin{equation}\label{eq:paramVariation}
    \hat{M}_x^{(i)} \to \hat{M}_x^{(i)} + \delta_x^{(i)} {{{{}\hat{M}}_x}^{(i)\bot}}.
    \end{equation}
    \item Solve the first order equations
    \begin{equation}\label{eq:firstOrderVariation}
    \bra{\psi} \frac{\partial \hat S}{\partial \delta_x^{(i)} } \ket{\psi}=0\ \ \forall x,i.
    \end{equation}
\end{enumerate}

Note that in Step 2 the measurements should be chosen such that $\text{span}\{\hat\Pi_{a_1|x_1}^{(1)}\otimes\dots \otimes \hat\Pi_{a_n|x_n}^{(n)}\}$ contains the Bell operator $\hat S$. Furthermore, when the measurement operators for at least one party define an overcomplete operator basis (not counting the identity), several choices of $\beta$ could be made in Step 3. In this case, the derivative in \cref{eq:firstOrderVariation} is to be understood accordingly.

Eigenvalue perturbation implies that \cref{eq:firstOrderVariation} must be satisfied when $\beta$ is maximized by the considered state and settings. Importantly, the method gives a \emph{necessary condition} for the Bell expression candidate $\beta$ obtained by the choice of operator $\Hat S$ and of settings $\Hat M_x^{(i)}$ that was made in the first place. 

Note that expression $\beta$ may still be maximally violated by the considered state even when condition \cref{eq:firstOrderVariation} is not verified. This is however only possible with other measurement settings and thus a different corresponding operator $\Hat S'$. For example, we showed previously that Bell expression~\cref{eq:bellcandidatenotgood} is not a ``good'' candidate for the operator choice $\Hat S = (\hat X_A \hat X_B + \hat Z_A \hat Z_B)/2$. Nevertheless, its maximum $2/\sqrt{3} \simeq 1.15>1$ is attained by the state $\ket{\phi^+}$ for the different measurement settings $a_1=0$, $a_2=\pi/2$, $b_y = -(-1)^y \pi/3$, and \cref{eq:firstOrderVariation} is satisfied for the corresponding Bell operator. It may thus be helpful to consider several Bell operators $\Hat S$.

The operator $\hat S$ belongs to a finite dimensional product Hilbert space and therefore can in principle be fully parametrized. The constraints here are that the state $\ket{\psi}$ should belong to the support of the considerd Hilbert space and be a maximal eigenstate of $\hat S$. Furthermore, finite dimensional measurements can also be expressed with a finite number of parameters. In all generality, the whole variational method can therefore be parametrized with finitely many parameters. When considering a complete parametrization rather than a particular choice of $\Hat S$ and of the measurements, the implications of \cref{eq:firstOrderVariation} become stronger.

First, if an expression $\beta$ is never obtained for all possible $\Hat S$ and $\Hat M_x^{(i)}$ verifying \cref{eq:firstOrderVariation}, then this expression cannot be maximized by the considered state. This implies that this expression cannot be used to self-test the considered state. When considering a complete parametrization, the variational method can thus be considered as a fully necessary condition for both Bell expression maximization and self-testing~\footnote{Note that when using the variational method to construct self-testing candidates, the operator $\hat S$ can be further restrained to admit the target state $\ket{\psi}$ as a unique maximum eigenstate.}.

Second, if a choice of measurements never verifies \cref{eq:firstOrderVariation} for all possible $\Hat S$, then these settings cannot be used in any expression maximized by the considered state. In particular, these settings cannot be used to self-test the considered state, or in other words, the realisation corresponding to the considered state and those measurements cannot be self-tested. \\

From the previous example, we note that an easy way to choose the initial Bell operator $S$ is to identify a set of operators $S_i$ that are stabilizers of the target state: $S_i\ket{\psi}=\ket{\psi}$ and such that $S_i$ have eigenvalues in $[-1,1]$. Then any convex sum of several of these $S = \sum_i p_i S_i$ with $p_i\geq0$ and $\sum_i p_i=1$ results in a valid operator $S$. 

As demonstrated in previous works, the variational method allows one to construct insightful Bell expressions tailored to various states. In~\cite{Sekatski18}, the method is used to find an expression self-testing the four qubits linear cluster state. The method also provided a self-test more robust to noise than the tilted CHSH inequality for the partially entangled two-qubit states~\cite{Wagner20} through
\begin{equation} \label{eq:Sebineq}
\begin{split}
\frac{\langle A_1B_1\rangle+\langle A_1B_2\rangle}{2\cos(b_\theta)} + s_{2\theta}\frac{\langle A_2B_1\rangle-\langle A_2B_2\rangle}{2\sin(b_\theta)}&\\
\quad \quad \quad \quad \quad \quad + \frac{1}{2}c_{2\theta}\left(\langle A_1\rangle + \frac{\langle B_1\rangle+\langle B_2\rangle}{2\cos(b_\theta)}\right) &\preceq 2,
\end{split}
\end{equation}
where $b_\theta = \pi/2 - \text{arctan}\sqrt{\frac{1+\frac{1}{2}c_{2\theta}^2}{s_{2\theta}^2}}$. The associated measurement settings are given by\footnote{Note that we changed the parametrization of the measurement angles compared to what is presented in~\cite{Wagner20} in order to be keep coherence throughout this article.}:
\begin{equation}
    \left\{\begin{split}
        & \Hat M_1^{(1)} = \Hat Z_A, \quad M_1^{(1)} = \Hat X_A, \\
        & \Hat M_y^{(2)} = \cos(b_\theta)\Hat Z_B-(-1)^y\sin(b_\theta)\Hat X_B.
    \end{split} \right.
\end{equation}

\subsubsection{Second order condition}

For an expression to be a maximum, its first derivative must be zero, but the second derivative should also be negative. The variational approach can thus be pushed one step further. As we show below, the negativity of the second derivative is often verified. Furthermore, in the case of multi-dimensional variation, the associated Hessian provides interesting insight about the interplay between various measurement parameters.

For a perturbation of the operator $\hat S$ induced by an infinitesimal change of the measurement settings in the Bell expression accordingly to \cref{eq:paramVariation}, the eigenvalue $1$ of $\ket{\psi}$ is a local maximum iff (in addition to \cref{eq:firstOrderVariation}) the following semi-definite condition holds:
\begin{subequations}
    \begin{align}
        &\gamma \preceq 0, \ \text{where } \ \gamma = \mu + \nu \\
        &\mu_{ij} = \bra{\psi} \frac{\partial \hat S}{\partial \delta_i \partial \delta_j} \ket{\psi} \\
        &\nu_{ij} = 2 \sum_l \frac{\bra{\psi} \frac{\partial \hat S}{\partial \delta_i} \ket{\psi_l}\bra{\psi_l} \frac{\partial \hat S}{\partial \delta_j} \ket{\psi}}{1-\lambda_l}.
    \end{align}
\end{subequations}
Here, $\ket{\psi_l}$ are the other eigenstates of $\hat S$ with the eigenvalues $\lambda_l$, $\mu$ accounts for the direct second order variation of the Bell operator on $\ket{\psi}$, while $\nu$ accounts for the variation of the eigenstate associated to the maximal eigenvalue~\cite{eigenvalueperturbation}.

As an example, consider the case of the Bell operator \cref{eq:XXZZOperator} with measurements \cref{eq:XZmeas}. The second order perturbation of the measurements is given by
\begin{equation}
\begin{split}
\hat A_x &\to \hat A_x + \delta_{a_x}(-\sin(a_x) \hat Z_A + \cos(a_x) \hat X_A) - \frac{1}{2}\delta_{a_x}^2 \hat A_x, \\
\hat B_y &\to \hat B_y + \delta_{b_y}(-\sin(b_y) \hat Z_B + \sin(b_y) \hat X_B) - \frac{1}{2}\delta_{b_y}^2 \hat B_y. 
\end{split}
\end{equation}
Considering only variations of the second measurement of each party (by invariance under local unitaries), we can impose the first order conditions and compute the Hessian matrix: 
\begin{equation}
\gamma= \begin{pmatrix}
-1 & \cos(2c)\\
\cos(2c) & -1
\end{pmatrix}.
\end{equation}
Once again the second order condition only depends on the single parameter $c=b_1-a_1$. The eigenvalues of $\gamma$ are $\{-2\sin(c)^2,-2\cos(c)^2$\}. For $c\in (0,\pi/4]$ those are strictly negative and are equal when $c=\pi/4$. These eigenvalues describe the decrease of the maximal value of the Bell operators to second order along principal measurements perturbation directions. When both eigenvalues are equal, we recover the CHSH expression. One can thus understand CHSH as the solution among all expressions in \cref{eq:betac} which behaves the most uniformly with respect to measurement perturbations.

\subsubsection{Limitations of the variational method}

As mentioned earlier, the variational method often allows one to construct Bell expressions which not only have the desired maximization property, but sometimes also achieve self-testing of the target state. However, the method only provides in general a necessary condition for a Bell inequality to be maximally violated by the target state. Indeed, since it only focuses on local extrema, it may fail to provide Bell expressions with matching global Tsireslon bound: the local maximum may fail to be a global one. In this section, we provide an example where the maximum obtained by the variational method is local but not global, thus showing that this is indeed a limitation of the method.

Concretely, we apply the method to the partially entangled two qubit state $\ket{\phi_\theta} = c_\theta \ket{00} + s_\theta \ket{11}$ for $\theta \in (0,\pi/4]$ using the Bell operator
\begin{equation}
\begin{split}
    \hat S_{\theta,p,q} = & p \hat Z_A \hat Z_B \\
    & + (1-p)(s_{2\theta} \hat X_A \hat X_B + c_{2\theta} (q\hat Z_A+(1-q)\hat Z_B))
\end{split}
\end{equation}
obtained by combining two stabilizers. This ensures that the Bell operator satisfies $\hat S_{\theta,p,q} \ket{\phi_\theta}=\ket{\phi_\theta}$. In addition, $\ket{\phi_\theta}$ is the unique maximal eigenvector of $\hat S_{\theta,p,q}$ when
\begin{equation}\label{eq: maxeig}
4p + (1-p)^2 2 q(1-q) ( \cos (4 \theta )+1)> 0.
\end{equation}

Next we consider arbitrary measurements in the $\hat X$-$\hat Z$ plane parameterized as in \cref{eq:XZmeas}. The first order conditions are: 
\begin{equation}
    \begin{split}
        ps_{a_2}s_{a_1} + (1-p)(s_{2\theta}^2c_{a_2}c_{a_1}+qc_{2\theta}^2 s_{a_2}s_{a_1}) &= 0, \\
        ps_{b_2}s_{b_1} + (1-p)(s_{2\theta}^2c_{b_2}c_{b_1}+(1-q)c_{2\theta}^2 s_{b_2}s_{b_1}) &= 0.
    \end{split}
\end{equation}
For concreteness, let us now set $\theta=\pi/8, a_1 = 0, b_1 = -b_2 = \pi/6$. The values of $a_2$ is fixed by the first order conditions to be $a_2=\pi/2$, and the second equation fixes the value of $p$ to
\begin{equation}
    p(q) = \frac{2+q}{4+q}.
\end{equation}
\cref{eq: maxeig} is then fulfilled if $q\in[0,4]$. This fixes the candidate Bell expression to
\begin{equation}
\scalebox{0.9}{$\begin{aligned}
    \beta_q &=  p(q) \frac{\langle A_1B_1\rangle+\langle A_1B_2\rangle}{\sqrt{3}}  + (1-p(q))\\
    &\times  \left[\frac{\langle A_2B_1\rangle-\langle A_2B_2\rangle}{\sqrt{2}}+ \frac{q\langle A_1\rangle}{\sqrt{2}} + (1-q)\frac{\langle B_1\rangle+\langle B_2\rangle}{\sqrt{6}}\right].
\end{aligned}$}
\end{equation}
We still need to check the second order condition to make sure that the value 1 is a local maximum. The non-zero eigenvalues of the Hessian matrix can be computed numerically and are all negative within the range of $q\in [0,4]$, thus guaranteeing that the Bell expression corresponds to a local maximum.

\begin{figure}
\centering
    \includegraphics[scale=0.6]{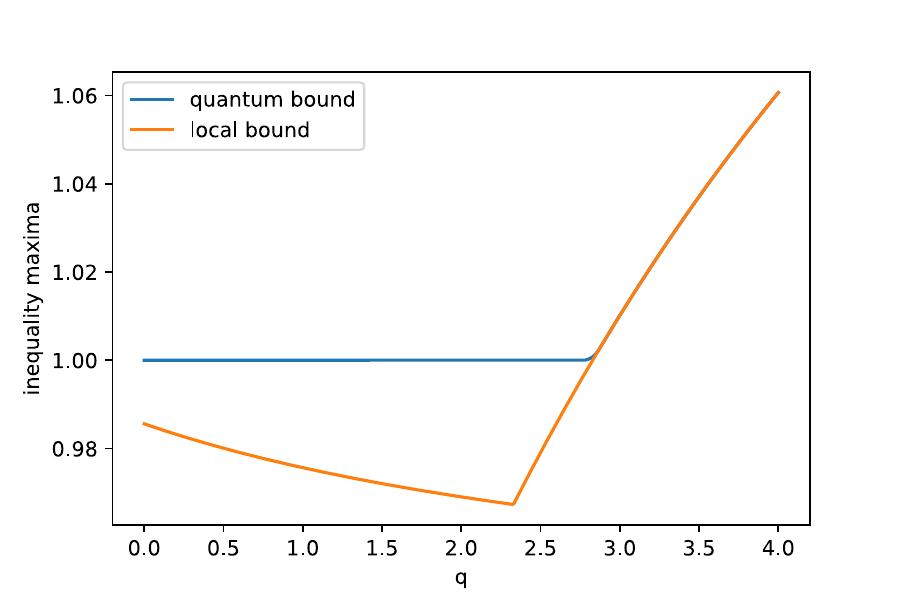}
    \caption{In blue, the numerical value of the quantum bound of our candidate Bell expression using semi-definite programming at the order 1+AB of the NPA hierarchy~\cite{Navascues08}. In orange, the numerical value of the local bound of our candidate Bell expression. For our settings choice, the value of the expression is $1$ for all values of $q$. The blue curve is equal to $1$ only in the region (0,2.83).}
    \label{fig:quantumphitheta2}
\end{figure}

In order to assess whether this local maximum is also global, we compute the quantum maxima numerically (see~\cref{fig:quantumphitheta2}).
As we can see the candidate Bell expression admits a quantum maxima strictly greater than $1$ when $q>2.83$. More than this, its local maxima is also larger than $1$: local realizations far away from the point of study can reach a larger value of the expression. One such point is given by the deterministic strategy $A_x = 1$, $B_y=-(-1)^y$ which gives $\beta_3(\bm P_L) = 1.01 >1$. Only in a smaller region of parameter $q$, approximately in $(0,2.83)$, the variational method seems to give Bell expressions with the expected maximal value.

\subsection{Sum of squares decomposition method}\label{sec:SOSmethod}

The variational method constructs Bell expression that are potentially maximized by a target quantum state, but it is fundamentally unable to guarantee that no other quantum realization could achieve a higher score. We now present a method that is also able to provide Bell expression tailored to a target state with the guarantee that no other quantum state can attain a higher score.

\subsubsection{Formal polynomials}

Let $\{X_i\}_i$ be a set of indeterminates in an associative algebra over a field $\mathbb{K}$. A \textit{formal multivariate polynomial} is a linear combination~\cite{Pironio10b}
\begin{equation}
S=\sum_i \alpha_i M_i
\end{equation}
where $M_i$ are monomials, i.e.~products of indeterminates such as $X_1$, $X_1X_2$ or $X_1^2 X_2 X_1$, weighted by scalars $\alpha_i\in\mathbb{K}$.

Considering the algebra induced by the Bell scenario, we associate to the outcome $a$ of the measurement $x$ of the party $k$ the indeterminate $X_{a|x}^{(k)}$. These indeterminates, also called `non-commuting variables', obey the algebraic rules of
\begin{subequations}
\begin{align}
    \text{Hermiticity:} &&\left(X_{a|x}^{(k)} \right)^\dagger &= X_{a|x}^{(k)}\\
    \text{Orthogonality:} &&X_{a|x}^{(k)}X_{a'|x}^{(k)} &= \delta_{a,a'} X_{a|x}^{(k)}\\
    \text{Normalization:} &&\sum_a X_{a|x}^{(k)} &= 1\\
    \text{Commutation:} &&[X_{a|x}^{(k)},X_{a'|x'}^{(k')}] &= 0\text{ for }k\neq k'.
\end{align}
\end{subequations}
In this algebra, any monomial $M_i$ can always be written as the product of $n$ multivariate monomials $M_i^{(k)}$ involving indeterminates from each party individually:
\begin{equation}
    M_i[\{X_{a|x}^{(k)}\}]= M_i^{(1)}[\{X_{a|x}^{(1)}\}]...M_i^{(n)}[\{X_{a|x}^{(n)}\}].
\end{equation}

When all monomials $M_i^{(k)}$ in the polynomial $S$ are of degree one, we say that $S$ is of \textit{local degree} 1~\cite{Moroder13}. We then associate the Bell expression
\begin{equation}
    \beta = \sum_{\bm a|\bm x}\alpha_{\bm a|\bm x} P(\bm a|\bm x)
\end{equation}
to the polynomial by substituting each indeterminate $X_{a|x}^{(k)}$ by the projector $\hat \Pi_{a|x}^{(k)}$ associated to the outcome $a$ for the setting $x$, and considering the expectation value over a state $\ket{\psi}$. This is always possible because an arbitrary implementation can be defined by a set of projective measurements $\{\hat \Pi_{a_k|x_k}^{(k)}\}$ and a pure state $\ket{\psi}$ (upon increasing the dimension of the local Hilbert spaces), giving the probabilities $P(\bm a|\bm x) = \bra{\psi} \bigotimes_k \hat \Pi_{a_k|x_k}^{(k)} \ket{\psi}$. Furthermore, this identification is reversible, defining for each Bell expression a corresponding unique formal polynomial. We thus refer to polynomials $S$ with local degree 1 as \textit{formal Bell polynomials}. As an example, the formal CHSH polynomial associated to the Bell expression defined in \cref{eq:CHSHBellExpression} is given by
\begin{equation}
    S_{\text{CHSH}} = \!\!\!\!\!\!\sum_{a_1,a_2,x_1,x_2}\!\! \!\!\!\!(-1)^{a_1+a_2+(x_1-1)(x_2-1)} X_{a_1|x_1}^{(1)} X_{a_2|x_2}^{(2)}.
\end{equation}

Any Bell expression can then be understood as the expectation value of a formal Bell polynomial by defining the expectation values of indeterminate monomials by the moments of unknown operators on an unknown state:
\begin{align}
\langle M_i \rangle &= \bra{\psi}\hat M_i\ket{\psi},
\end{align}
where $M_i$ is an implementation of the operators on some Hilbert space, and $\ket{\psi}$ is a state in this space. When the local degree of the monomial $M_i$ is one, this moment admits a decomposition in terms of conditional probabilities $P(ab|xy)$, see \cref{eq:CorrelatorsInProbabilities}.

Note that the relation between Bell expressions and operators described in~\cref{sec:generalDefs} also associates to every formal Bell polynomial a Bell operator. The substitution of $X_{a|x}^{(k)}$ by the projector $\hat \Pi_{a|x}^{(k)}$ extends this identification to formal polynomials beyond local degree 1.

Just like it is sometimes more convenient to parametrize measurements by the combination of projectors, it is sometimes useful to represent a formal polynomial in terms of different indeterminates than the ones associated with measurement projection operators. For instance, each measurement $x$ of party $k$ can be described by the single operator
\begin{equation}\label{eq: FT measuremnt opearators}
     Y_{x}^{(k)} = \sum_{a=0}^{d-1} w_d^a X_{a|x}^{(k)},
\end{equation}
where $w_d=\exp(2i\pi/d)$, obtained by applying the Fourier transform on all indeterminates associated with the different outcomes of the measurement. By construction, the d-th power of $Y_{x}^{(k)}$ is $1$, and all the indeterminates $X_{a|x}^{(k)}$ can be expressed as a linear combination of the power of $Y_{x}^{(k)}$. Note that this requires considering a complex field $\mathbb{K}$ when $k>2$. The new indeterminates $Y_{x}^{(k)}$ are unitary in the sense that $\left(Y_x^{(k)}\right)^\dag Y_x^{(k)} = Y_x^{(k)} \left(Y_x^{(k)}\right)^\dag = 1$. In this new basis, the CHSH polynomial takes the form:
\begin{equation}
    S_{\text{CHSH}} = A_1B_1 + A_1B_2  + A_2B_1  - A_2B_2,
\end{equation}
where we used a more elegant notation for the indeterminates $A_x=Y_x^{(0)}$ and $B_y=Y_y^{(1)}$ in the bipartite case. Note that the Fourier indeterminates are also Hermitian $A_x^\dag=A_x, B_x^\dag=B_x$ in the case of binary measurements $(d=2)$.

Since they provide a description of Bell expressions independently of a specific quantum implementations, formal polynomials on non-commuting variables provide a powerful tool for their analysis. In particular, formal polynomials can be used to compute Tsirelson bounds on Bell expressions, i.e.~bounds on the maximal quantum value achievable that is valid independently of the quantum state, measurement and Hilbert space dimension.

As an example, it is known that the CHSH polynomial can be written as the following sum of squares~\cite{Doherty08,Reichardt13,Bamps15,Supic20}:
\begin{equation}
\begin{split}
    2&\sqrt{2} - S_{\text{CHSH}} = \\
    &\ \ \frac{1}{\sqrt{2}} \left[ \left( A_1 - \frac{B_1+B_2}{\sqrt{2}} \right)^2 + \left( A_2 - \frac{B_1-B_2}{\sqrt{2}} \right)^2\right].
\end{split}
\end{equation}
Since the square of a polynomial $S^\dag S$ can only result in a positive contribution, this directly implies that the maximal value of the corresponding CHSH expression achievable upon measurement of a quantum state is limited to $2\sqrt{2}$. In other words, polynomial inequalities
\begin{equation}
    S \preceq C
\end{equation}
on formal Bell polynomials $S$, where $C$ is a real number, imply a Tsireslon bound on the corresponding Bell expression: for all possible statistics $\bm P$ obtained by quantum mechanics, the associated Bell expression $\beta$ verifies
\begin{equation}
\beta(\bm P) \leq C.
\end{equation}

\subsubsection{The SOS method}
Suppose that a formal polynomial of local degree $1$ can be decomposed as a sum of squares, \textit{i.e}: 
\begin{equation}
    S = C - \sum_i N_i^\dag N_i,
\end{equation}
where $N_i$ are formal polynomials and $C$ is a real number. For every implementation, we have $\bra{\psi} \hat N_i^\dag \hat N_i \ket{\psi} \geq 0$, where $\hat N_i$ is the Bell operator associated to $N_i$ by the implementation. Thus the maximal quantum score of the Bell expression over any implementation is upper bounded by $C$. If a specific implementation gives exactly $C$ then we get $\sum_i ||\hat N_i\ket{\psi}||^2 = 0 \Rightarrow \ \forall i, \, ||\hat N_i\ket{\psi}||^2 = 0$ and $\ket{\psi}$ is in the kernel of all  operators $\hat N_i$. When this is the case, we say that the operators nullify the state. The opposite is also true, hence
\begin{equation}
    \bra{\psi} \hat S\ket{\psi} = C \ \iff  \ \forall\, i, \ \hat N_i \ket{\psi} =  0.
\end{equation}
In summary, being able to write a Bell polynomial as a sum of squares allows one to get an upper bound on the maximal score of the corresponding Bell expression~\cite{Doherty08}. Moreover, if all squares nullify the target state for some choice of measurements, then the upper bound is strict and is obtained by this implementation.

However, in general for arbitrary polynomials $N_i$, a sum of squares gives
\begin{equation}
    \sum_{i\in I} N_i^\dag N_i = C - S + \Gamma,
    \label{sos-t}
\end{equation}
where $C$ is a real number, $S$ is a formal Bell polynomial (to which we can associate a Bell expression), and $\Gamma$ is some leftover polynomial term of higher local order. This last term needs to vanish in order to have a valid sum of squares decomposition. The ``condition'' of the sum of squares method thus take the simple form
\begin{equation}\label{eq:SOScondition}
\Gamma=0.
\end{equation}

Building on the variational method, where the measurement parameters (and thus Bell coefficients) are chosen in order to satisfy some local extremality condition, we propose to set these parameters in such a way that the condition $\Gamma=0$ holds. This leads us to formulate the following SOS method for the construction of Bell expressions.
\begin{enumerate}
    \item 
    Choose a set of operators $\hat N_i$ that are nullifying the target state, i.e.~such that $\ket{\psi}\in \cap_i \ker(\hat N_i)$, and in particular
    \begin{equation}
        \hat N_i \ket{\psi}=0\ \ \forall i.
    \end{equation}
    \item 
    Parametrize measurement bases $\hat M_{x}^{(i)}$ for each party.
    \item
    Express the nullifiers in terms of the measurement operators $\hat M_x^{(i)}$ and define each of the corresponding formal polynomials by promoting the measurement operators to indeterminates.    \item 
    Compute the sum of squares \cref{sos-t} on the polynomials $N_i$.
    \item
    Solve the condition that all terms of local order higher than one vanishes
    \begin{equation}
        \Gamma = 0.
    \end{equation}
\end{enumerate}

As an illustration, consider the following application of the SOS method on the singlet state\footnote{Since we are only interested in identifying states up to local unitaries here and below we slightly abuse the terminology and refer to any maximally entangled two qubit state as the singlet state.} $\ket{\psi}=\ket{\phi^+}$. A simple choice of nullifiers is given by $\hat N_0=\hat Z_A-\hat Z_B$ and $\hat N_1=\lambda(\hat X_A-\hat X_B)$, with $\lambda\in\mathbb{R}$. We can then make a simple choice of measurement by choosing $\hat M_1^{(1)}=\hat Z_A$, $\hat M_2^{(1)}=\hat X_A$, $\hat M_y^{(2)} = \cos(b) \hat Z_B - (-1)^y \sin(b) \hat X_B$, where $b\in[0,\pi/2]$ is an arbitrary angle. This allows us to re-express the nullifiers as $\hat N_0=\hat M_1^{(1)}-\frac{\hat M_1^{(2)}+\hat M_2^{(2)}}{2\cos(b)}$, $\hat N_1=\lambda\big(\hat M_2^{(1)}-\frac{\hat M_1^{(2)}-\hat M_2^{(2)}}{2\sin(b)}\big)$. Substituting the $\hat M_x^{(i)}$ operators with the indeterminates $Y_x^{(i)}$, we express their sum of squares as
\begin{equation}
\begin{split}
    N_0^2 + N_1^2 =& 1 + \frac{1}{2\cos^2(b)} + \lambda^2\left(1+\frac{1}{2\sin^2(b)}\right)\\
    &- \left( A_1\frac{B_1+B_2}{\cos(b)} + \lambda^2 A_2\frac{B_1-B_2}{\sin(b)} \right)\\
    & + \frac{1}{4}\left( \frac{1}{\cos^2(b)} - \frac{\lambda^2}{\sin^2(b)}\right)\{B_1,B_2\}.
\end{split}
\end{equation}
The term containing $\{B_1,B_2\}$ is the only one with local order higher than $1$. It vanishes for the choice $\lambda^2 = \tan^2(b)$, yielding $N_0^2+N_1^1=C(b)-S(b)$ with the Bell expression
\begin{equation}
    S(b) = \frac{A_1(B_1+B_2) + \tan(b)A_1(B_1-B_2)}{\cos(b)}
\end{equation}
and the Tsirelson bound $S(b)\preceq C(b)= 2(1+\tan^2(b))$ attained by the target state $\ket{\phi^+}$. One can check that achieving this value self-tests the state.\\

The promotion of the nullifiers from the operator space to the formal polynomial algebra performed in Step 3 before computing the sum of squares is a key ingredient of the SOS method. While the choice space of the nullifiers $\hat N_i$ only depends on the target state, this mapping from $\hat N_i$ to $N_i$ depends on the choice of measurements. When evaluated on specific measurements, these polynomials correspond to operators that nullify the target state. We thus refer to them as \emph{formal nullifiers} for the considered state and measurements.

Note that as in the variational method, when using the SOS method to obtain a self-test candidate, rather than simply a Bell expression maximized by the target state, one may want to choose nullifiers in such a way that the target state is the unique one nullified by all of them, i.e.~$\cap_i \ker(\hat N_i) =\text{span}\{\ket{\psi}\}$. Indeed, consider the particular case $\lambda=0$ to the above example: we can expand the unique nullifier as $\hat N_0 =\hat Z_A-\hat Z_B$, which yields the SOS decomposition $N_0^2 = (A_1-B_1)^2= 2-S$ with the candidate Bell expression $S=A_1B_1$. This successfully yields the candidate $S=2A_1B_1$, along with SOS decomposition $2-S=(A_1-B_1)^2$. But since only one square appears in the decomposition, only the nullifying equation $(\hat M_1^{(1)}-\hat M_1^{(2)})\ket{\psi}=0$ is certified when the quantum value $2$ is reached. Clearly, this is not sufficient to identify the state uniquely since even in the ideal implementation there are many candidates that can achieve the maximal value of this expression: many states are nullified by $\hat Z_A-\hat Z_B$; for instance $\ket{\phi^+}$ but also $\ket{00}$ and $\ket{11}$.

The SOS method guarantees that no implementation can provide a larger Bell score to the Bell expression $\beta$ than the target one. It is thus a \emph{sufficient condition} to construct a Bell expression for a target quantum state: any expression verifying $\Gamma=0$ is maximized by the considered state. The method relies on a choice of operators, here nullifiers, and of measurements. Therefore, a candidate that does not verify the SOS condition~\cref{eq:SOScondition} for a specific choice of $\Hat N_i$ and $\Hat M_x^{(i)}$ is not automatically ruled out as it might admit a valid SOS decomposition for another choice. This also means that a parametrization of the measurements in Step 2 need not be complete in order to apply the method, which may be challenging to achieve for high dimensional target Hilbert spaces, see also \cref{sec:qutrits}.

Note that the nullifiers chosen in Step 1 act on a finite dimension Hilbert space and so they can be fully parametrized. However, their number is not bounded (they can be linearly dependent). Moreover, each nullifier can be expressed in terms of the measurement operators in numerous ways. In particular, their expressions do not need to be restricted to local degree one. Indeed, the space of polynomials corresponding to a formal nullifier for a given state and measurements is of infinite dimension. Therefore, formal nullifiers cannot be parameterized with a bounded number of parameters in full generality. The number of parameters is however finite when the length of the monomials that can be used in the formal polynomials is bounded.

It is known that finding the SOS decomposition of a Bell expression~\cite{Doherty08} is dual to the NPA hierarchy~\cite{Navascues08}, which converges in the limit of the hierarchy. This limit corresponds to considering the full space of formal polynomials with no limit on the number of squares and on the monomial length. Thus, if one considers all possible choices of formal nullifiers for monomial length $n$, and completely parametrizes the finite dimensional measurements, the SOS method becomes \emph{necessary and sufficient} in the asymptotic limit $n\to \infty$. This means that a Bell expression can be maximized by the considered state if and only if there exists a choice of measurements and an asymptotic choice of nullifiers such that $\Gamma=0$. In this sense, the SOS method is \emph{asymptotically complete}.

Note that contrary to the variational method which cannot ensure that its maxima is global, the SOS method provides this guarantee. Moreover, since a global maximum is a local one as well, its solution always fulfills the variational condition. Indeed, the SOS condition $C-S = \sum_i N_i^2$ is verified for the target implementation, where it gives the opertor equality $C\id - \Hat{S} = \sum_i \Hat{N}_i^2$. 
Since $\Hat{S} \ket{\psi} = C \ket{\psi}$, $\Hat{S}$ defines a valid Bell operator for the variational method. The conditions of the variational method are then satisfied: the target state is an eigenvector of maximal eigenvalue and all first order equations upon variations of the measurements choices vanish. This observation can be used to restrict the choice of nullifiers and/or of measurements settings that one can use in the SOS method by first applying the variational method.

Finally, we remark that the strength of this method is not only to construct Bell expressions whose maximal value is reached by the target state. By also providing their sum of squares decomposition, conditions on the action of the measurements on the state are also obtained: the operators $\Hat N_i$ nullify the state for any implementation reaching the quantum bound. This is interesting as for an arbitrary Bell expression, it is in general hard to find its quantum bound or its SOS decomposition. Moreover, many proofs of self-testing rely on the sum of squares decomposition of the Bell test \cite{Supic20}. We discuss this point in more details in~\cref{sec:analyticSelftest}.

\section{Applications and results}

Here we present applications of the SOS method to several states and Hilbert spaces, demonstrating how it may be used to derive Bell inequalities maximized by target states and self-test for a variety of cases.

\subsection{Recovering all self-tests of the singlet with two binary measurements as linear self-tests}

In this section, we use the SOS method to derive Bell expressions that self-test the singlet for all possible measurement settings. We apply the SOS method to the state $\ket{\phi^+} = (\ket{00}+\ket{11})/\sqrt{2}$ allowing operators to be taken in $\langle \{1,\hat Z,\hat X\}^{\otimes 2} \rangle$.

Consider a subspace of nullifiers given by:
\begin{equation*}
    \mathcal{A} = \langle \{ \Hat Z_A - \Hat Z_B , \Hat X_A -\Hat X_B \} \rangle.
\end{equation*}
The only two qubit state nullified by any two linearly independent elements of $\mathcal{A}$ is the singlet. Therefore a good candidate for nullifiers $\Hat N_i$ would be to take two such operators $\Hat N_0,\Hat N_1 \in \mathcal{A}$:
\begin{equation}
    \left\{ \begin{split}
        & \Hat N_0 = \alpha (\Hat Z_A - \Hat Z_B) + \beta (\Hat X_A - \Hat X_B),\\
        & \Hat N_1 = \gamma (\Hat Z_A - \Hat Z_B) + \delta (\Hat X_A - \Hat X_B),
    \end{split}\right.
    \label{trick}
\end{equation}
where $\alpha,\beta,\gamma,\delta$ are arbitrary real numbers.

Consider measurement setting in the $\Hat X$-$\Hat Z$ plane for both Alice and Bob. Up to local unitaries and relabeling of outcomes we can always write
\begin{equation}
\begin{split}
    &\left\{ \begin{split}
        & \Hat M_1^{(1)} = \Hat Z_A, \\
        & \Hat M_2^{(1)} = \cos(a_2) \Hat Z_A + \sin(a_2) \Hat X_A,
    \end{split}\right.\\
    &\left\{ \begin{split}
        & \Hat M_1^{(2)} = \cos(b_1) \Hat Z_B + \sin(b_1) \Hat X_B, \\
        & \Hat M_2^{(2)} = \cos(b_2) \Hat Z_B + \sin(b_2) \Hat X_B
    \end{split}\right.
    \label{measurements}
\end{split}
\end{equation}
with angles $a_2,b_1,b_2 \in [0,\pi[$. We can also assume up to relabeling of measurements that $b_1\leq b_2$.
In the basis of the measurements, we obtain
\begin{equation}\label{eq:SingletPaulis}
\begin{split}
    & \Hat Z_A=\Hat M_1^{(1)}, \quad \Hat X_A = \frac{\Hat M_2^{(1)} - \cos(a_2) \Hat M_1^{(1)}}{\sin(a_2)}, \\
    & \Hat Z_B= \frac{\sin(b_2)\Hat M_1^{(2)} - \sin(b_1)\Hat M_2^{(2)}}{\sin(b_2-b_1)}, \\
    & \Hat X_B= \frac{-\cos(b_2)\Hat M_1^{(2)} + \cos(b_1)\Hat M_2^{(2)}}{\sin(b_2-b_1)}.
\end{split}
\end{equation}

Replacing the operators $\Hat Z_A,\Hat X_A,\Hat Z_B,\Hat X_B$ with the above, we can express the nullifiers $\Hat N_i$ in terms of the measurements $\Hat M_x^{(i)}$. Let us now consider the formal polynomials associated to $\Hat N_0$ and $\Hat N_1$ and look at the sum of their squares
\begin{equation}
    N_0^2 + N_1^2 = C - S + \Gamma.
\end{equation}

The requirement $\Gamma=0$ for the measurement choice we made translates into two equations: 
\begin{equation}
    \left\{\begin{split}
        & \beta(\sin(a_2)\alpha-\cos(a_2)\beta)+\delta(\sin(a_2)\gamma-\cos(a_2)\delta)=0, \\
        & \begin{split}
            (\sin&(b_2)\alpha-\cos(b_2)\beta)(\sin(b_1)\alpha-\cos(b_1)\beta) \\
            &+(\sin(b_2)\gamma-\cos(b_2)\delta)(\sin(b_1)\gamma-\cos(b_1)\delta)=0.
        \end{split}
    \end{split} \right.
    \label{equationphi+}
\end{equation}
The first equation comes from cancelling the term containing $\{A_1,A_2\}$ in $\Gamma$, and the second from the $\{B_1,B_2\}$ term. We now look for a solution to this set of equations.

One solution is given by the choice $\alpha=1,\beta=0$, which gives
\begin{subequations}
    \begin{align}
        & \gamma = \frac{\cos(a_1)}{\sin(a_1)}\delta, \\
        & \delta^2 = \frac{1}{f(a_2,b_1,b_2)},
    \end{align}
\end{subequations}
where
\begin{equation}
f(a_1,b_0,b_1) = (\cot(a_2) - \cot(b_2))(\cot(b_1) - \cot(a_2)).
\end{equation}
This set of equation only admits solutions when $f(a_1,b_0,b_1)> 0$ which is the case when operators of Alice and Bob ``alternate'' stricly, \textit{i.e} $b_1< a_2<b_2$. We thus find a Bell expressions for all settings that can be self-tested for the singlet, as described in \cite{Wang16}, except for the ``limit points'' for which Alice and Bob share one common measurement. When this is the case, the initial set of equations, with no assumptions on $\alpha,\beta$, does admit a solution but the expressions we find are not satisfactory as the SOS decomposition only involves a single square. In the Appendix \ref{sec:Limitpoints}, we prove that all those limit points are non-exposed points of the set of quantum correlations and thus cannot be self-tested with a single Bell expression.

The sum of squares decomposition given by the method is
\begin{equation}
    N_0^2 + N_1^2 = C(a_1,b_0,b_1) - S(a_1,b_0,b_0)
\end{equation}
where
\begin{widetext}
\begin{subequations}
    \begin{align}
         N_0 &= A_1 - \frac{\sin(b_2)B_1 - \sin(b_1)B_2}{\sin(b_2-b_1)} \\
         N_1 &= \frac{1}{\sin(a_2)\sqrt{f(a_2,b_1,b_2)}} \left(A_2 - \frac{\sin(b_2-a_2)B_1 - \sin(b_1-a_2)B_2}{\sin(b_2-b_1)} \right) \\
         S(a_2,b_1,b_2) &= \frac{2}{\sin(b_2-b_1)}\left[ \sin(b_2)A_1B_1  + \frac{\sin(b_2-a_2)}{\sin^2(a_2)f(a_2,b_1,b_2)} A_2B_1 + \frac{\sin(a_2-b_1)}{\sin^2(a_2)f(a_2,b_1,b_2)} A_2B_2 - \sin(b_1)A_1B_2 \right] \label{bell_phi+S}\\
         C(a_2,b_1,b_2) &= \frac{2\sin(a_2)\sin(a_2-b_1-b_2)}{\sin(a_2-b_1)\sin(a_2-b_2)}.
    \end{align}
    \label{bell_phi+}
\end{subequations}
\end{widetext}
The Tsirelson bound associated to this sum of squares is given by
\begin{equation} \label{eq:allineqPhiP}
    S(a_2,b_1,b_2) \preceq C(a_2,b_1,b_2).
\end{equation}
We show in Appendix~\ref{sec:ProofPhiP} that all these Bell expressions indeed grant a self-test for the $\ket{\phi^+}$ state and their specific measurements settings.

Note that in \cite{Wang16}, the self-test of all those points is proven using the description of the border of the set of bipartite quantum correlation given by
\begin{equation}
\begin{split}
    \arcsin&(\langle  A_1B_1\rangle)+\arcsin(\langle A_2B_1\rangle) \\
    & +\arcsin(\langle A_2B_2\rangle)-\arcsin(\langle A_1B_2\rangle) = \pi
\end{split}
\end{equation}
By linearizing this equation for the border of this set we can obtain the equation of a tangent hyperplane to each point on the border. For the correlation point obtained with state $\ket{\phi^+}$ and previous measurements parameterized by $(a_2,b_1,b_2)$, we obtain the hyperplane
\begin{equation}
\begin{split}
    \mathcal{H}&_{a_2,b_1,b_2} = \big\{ (\langle A_1B_1\rangle,\langle  A_1B_2\rangle,\langle A_2B_1\rangle,\langle A_2B_2\rangle) \text{  s.t.} \\
    & \frac{1}{\sin(b_1)} \langle A_1B_1\rangle + \frac{1}{\sin(a_2-b_1)} \langle A_2B_1\rangle  \\
    & +  \frac{1}{\sin(b_2-a_2)} \langle A_2B_2\rangle - \frac{1}{\sin(b_2)} \langle A_1B_2\rangle \\
    & = \cot(b_1)+\cot(a_2-b_1)+\cot(b_2-a_2)-\cot(b_2) \big\}.
\end{split}
\end{equation}
One can check that up to re-normalization we can write the equation of the hyperplane as follows
\begin{equation}
    \begin{split}
    \mathcal{H}_{a_2,b_1,b_2} = \big\{ & (\langle A_1B_1\rangle,\langle  A_1B_2\rangle,\langle A_2B_1\rangle,\langle A_2B_2\rangle)  \\
    & \text{s.t.}\quad  S(a_2,b_1,b_2) = C(a_2,b_1,b_2) \big\} 
\end{split}
\end{equation}
Therefore the Tsirelson bounds we find match the equations of these hyperplanes at each point of the border of the quantum set.

\subsection{Self-tests for the partially entangled two-qubit states}

\subsubsection{A one-parameter family of self-test based on two nullifiers} \label{sec:partial1}
In this section, we look at partially entangled two qubit states
\begin{equation}
    \ket{\phi_\theta}=c_\theta\ket{00}+s_\theta\ket{11},
\end{equation}
where we use the notation $c_\theta=\cos(\theta)$, $s_\theta=\sin(\theta)$. We just slightly modify the nullifiers used for the singlet in the previous section to obtain the following nullifying operators
\begin{equation}
    \left\{ \begin{split}
        & \Hat N_0 = \Hat Z_A - \Hat Z_B, \\
        & \Hat N_1 = \Hat X_A - s_{2\theta}\Hat X_B - c_{2\theta}\Hat X_A\Hat Z_B
    \end{split}\right.
\end{equation}
for $\ket{\phi_\theta}$.

We parameterize the measurement operators $\hat M_x^{(i)}$ in the $\Hat X$-$\Hat Z$ plane with angles $a_x,b_y$ for Alice and Bob respectively as \cref{eq:XZmeas}. We then express the nullifiers in terms of the measurement operators, promote them to formal polynomials and introduce the formal sum of squares $N_0^2+\lambda^2 N_1^2$ for $\lambda \in \mathbb{R}$. The condition $\Gamma=0$ grants the following set of five equations:
\begin{equation}
    \left\{\begin{split}
        & s_{a_1}s_{a_2} + \lambda^2 \left(1+c_{2\theta}^2\frac{s_{b_1}^2 + s_{b_2}^2}{s_{b_1-b_2}^2}\right)c_{a_1}c_{a_2} =0, \\
        & s_{b_1}s_{b_2} + \lambda^2 \left(s_{2\theta}^2c_{b_1}c_{b_2} +c_{2\theta}^2\frac{c_{a_1}^2 + c_{a_2}^2}{s_{a_1-a_2}^2}s_{b_1}s_{b_2}\right) =0, \\
        & \lambda^2 c_{2\theta}c_{a_1}c_{a_2} =0,\\
        & \lambda^2 s_{2\theta}c_{2\theta}s_{b_1+b_2}=0,\\
        & \lambda^2 c_{2\theta}^2c_{a_1}c_{a_2}s_{b_1}s_{b_2}=0.
    \end{split} \right.
\end{equation}

For a fixed value of $\lambda$, this set of equations only admits one solution -- up to relabelling measurements and/or outcomes -- when $s_{a_1}=c_{a_2}=0$ and $b_1+b_2=0$. This means that the ideal measurements follow:
\begin{equation}
    \left\{\begin{split}
        & \Hat M_1^{(1)} = \Hat Z_A, \quad \Hat M_2^{(1)} = \Hat X_A, \\
        & \Hat M_y^{(2)} = \cos(b)\Hat Z_B-(-1)^y\sin(b)\Hat X_B,
    \end{split} \right.
    \label{meas_entangled}
\end{equation}
where the parameter $b$ satisfies
\begin{equation}
    \frac{1}{\lambda^2} = \sin^2(2\theta)\cot^2(b) - \cos^2(2\theta).
    \label{constrain}
\end{equation}
When this condition holds, the sum of squares gives a formal polynomial of local degree $1$ together with its SOS decomposition
\begin{equation}
    N_0^2 + \lambda^2 N_1^2 = C(\theta,b) - S_{\theta,b}
\end{equation}
where: 
\begin{equation} \label{bell_ineq}
    \scalebox{0.95}{$\begin{aligned}
        & N_0 = A_1 - \frac{B_1+B_2}{2\cos(b)} \\
        & N_1 = A_2 - s_{2\theta} \frac{B_1-B_2}{2\sin(b)} - c_{2\theta} A_2 \frac{B_1+B_2}{2\cos(b)} \\
        & S_{\theta,b} = A_1\frac{B_1+B_2}{\cos(b)} + \lambda^2 \left[s_{2\theta}A_2\frac{B_1-B_2}{\sin(b)} + c_{2\theta}\frac{B_1+B_2}{\cos(b)}\right] \\
        & C(\theta,b) = 2(1+\lambda^2)
    \end{aligned}$}
\end{equation}

Since $N_i\ket{\phi_\theta}=0$ for the considered implementation, the Tsirelson bound associated with the obtained Bell expression is 
\begin{equation}
    S_{\theta,b} \preceq 2(1+\lambda^2).
\end{equation}
But it turns out that we can obtain a stronger conclusion. In Appendix \ref{sec:ProofPartial} we prove that the maximal quantum value of this Bell expression self-tests the partially entangled state $\ket{\phi_\theta}$ and the measurements given in \cref{meas_entangled}.

Note that the left hand side of the \cref{constrain} is strictly positive. This implies a condition on the measurement angle $b$ and the entanglement parameter $\theta$ for a solution to exist. As such, we obtain: 
\begin{equation}
    b \in \left(\max(-2\theta,-\pi+2\theta),\min(2\theta,\pi-2\theta)\right)\, \setminus \, \{0\}
\end{equation}
It is an open question whether the limit points (for which $b = \min(2\theta,\pi-2\theta)$) can be self-tested or not. The method gives only unsatisfying candidates with decomposition into a single square. Our guess is that as for the singlet case, these points might self-test the underlying implementation but not with a single Bell expression -- \textit{i.e} they are non-exposed~\cite{Goh18,Chen23}.

One can study the second order of the variational method to choose a good candidate among all those self-tests. Considering only relative variations $\delta_{A/B} = \delta_{A_1/B_1} - \delta_{A_2/B_2}$ of Alice and Bob's measurement parameters, the Hessian matrix is: 
\begin{equation}
    \gamma_{\theta,b} = \begin{pmatrix}
    - \frac{\lambda^2 s_{2\theta}^2}{1+\lambda^2} & 0 \\
    0 & -\frac{1}{4}\left(1+\lambda^2 - \frac{\lambda^2 s_{4\theta}^2}{s_{2b}^2}\right)
    \end{pmatrix}
\end{equation}
Like in the singlet case we can look for settings for which the two eigenvalues are equals, so that the maximal eigenvalue of $\hat S$ drops equally for all nontrivial measurement perturbations. For $\theta \in (0,\pi/4]$, this is the case when $b=\theta$. In this case the two eigenvalues are equal to $-\sin^2(\theta)$.

In subsection \ref{sec:multipartite}, we generalize these expressions to an arbitrary number of parties $n$, by looking at partially entangled GHZ states, providing a first self-test of all states in this family in terms of a single Bell expression family.

\subsubsection{Insight on geometrical properties of the set of quantum correlations}
The Bell expressions given by \cref{bell_ineq} enable us to self-test states and settings for which self-testing was already known to be possible with other Bell expressions.  Indeed, the partially entangled states $\ket{\phi_\theta}$ can be self-tested using the so-called tilted CHSH inequality~\cite{Bamps15, Coopmans17} 
\begin{equation}
    I_{\alpha(\theta)} = S_{\text{CHSH}} + \alpha(\theta)  A_1 \preceq \sqrt{8+2\alpha^2},
\end{equation}
where $\alpha(\theta)=2/\sqrt{1+2\tan^2(2\theta)}$ for $\theta\in(0,\pi/4]$. The self-tested measurement settings are
\begin{equation}
    \left\{\begin{split}
        & \Hat M_1^{(1)} =\Hat Z_A, \quad \Hat M_2^{(1)} =\Hat X_A, \\
        & \Hat M_y^{(2)} = \cos(\mu_\theta)\Hat Z_B-(-1)^y\sin(\mu_\theta)\Hat X_B,
    \end{split} \right.
\end{equation}
where $\tan(\mu_\theta)=\sin(2\theta)$. Notice that $\mu_\theta\leq 2\theta$ and thus the correlation points achieved with the settings of tilted CHSH inequality can also be self-tested using \cref{bell_ineq} for the choice of $b\overset{!}{=} \mu_\theta$.

Since these two inequalities differ, the correlation points corresponding to the tilted CHSH expression can be self-tested using two different Bell expressions. In fact, it can be self-tested using any convex combination of the tilted CHSH inequality $I_{\alpha(\theta)}$ and the new Bell inequalities we presented $S_{\theta, \mu_\theta}$. Geometrically speaking, this means that this quantum point admits two tangent hyperplanes. This proves that the boundary of the quantum set admits non-local angulous point. This particular conclusion could also be inferred from the observation made in~\cite{Goh18} that some Bell inequalities are maximized by both the Tsirelson point and a local point.

\begin{figure}
    \centering
    \includegraphics[scale=0.55]{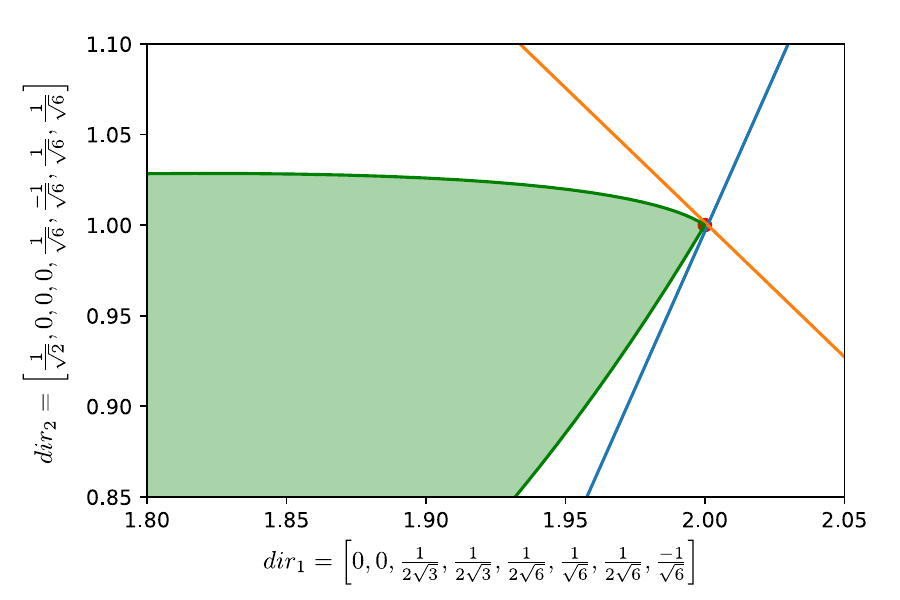}
    \caption{The green area shows the upper bound on the quantum set given by the NPA hierarchy at level 1+AB in the slice specified by $dir_1$ and $dir_2$. The directions are specified by $dir_x =[\langle A_1\rangle,\langle A_2\rangle,\langle B_1\rangle,\langle B_2\rangle,\langle A_1B_1\rangle,\langle A_2B_1\rangle,\langle A_1B_2\rangle,\langle A_2B_2\rangle]$. The red point is the quantum point achieved by the tilted CHSH settings for $\theta=\pi/8$. The orange line correspond to the tilted CHSH inequality at this point and the blue line correspond to our new Bell inequality $S_{\theta, \mu_\theta}$. The quantum set cannot go beyond those two lines.}
    \label{fig:angular}
    \centering
    \includegraphics[scale=0.55]{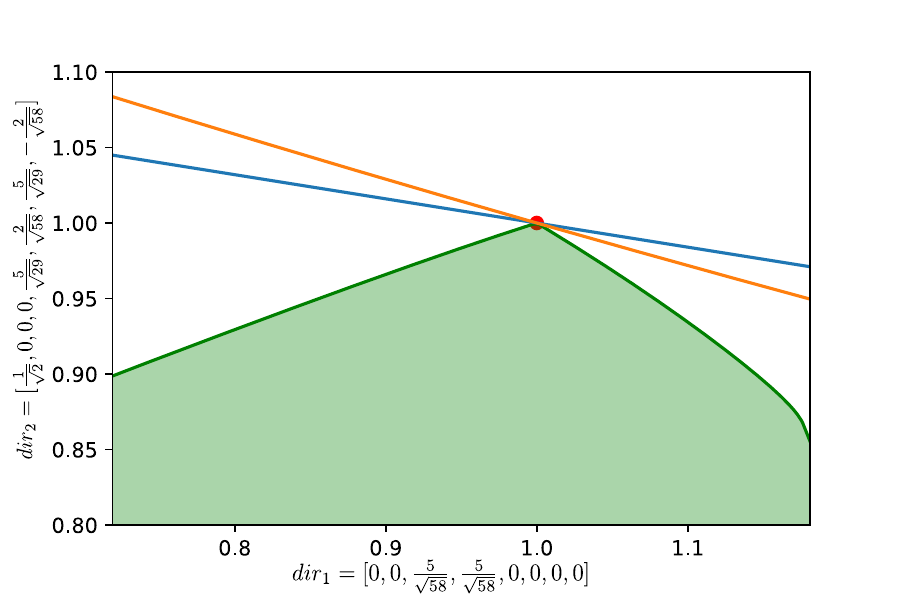}
    \caption{The green area shows the upper bound on the quantum set given by the NPA hierarchy at level 1+AB in the slice specified by $dir_1$ and $dir_2$. The red point is the quantum point achieved by the state and settings maximizing inequality \cref{eq:Sebineq} for $\theta=\pi/8$. The orange line correspond to this Bell inequality and the blue line correspond to our new Bell inequality $S_{\theta, b_\theta}$. The quantum set cannot go beyond those two lines.}
    \label{fig:angular2}
\end{figure}

The same analysis could be done for the points maximizing the expression \cref{eq:Sebineq}, as we again find novel Bell expressions for the same realisations. Those properties are illustrated in \cref{fig:angular} and \cref{fig:angular2}. 

\subsubsection{Self-testing the partially-entangled state when Alice's measurements are $\hat Z$ and $\hat X$}
The one-parameter family of Bell expressions \cref{bell_ineq} allows to self-test the state $\ket{\phi_\theta}$ with any $\theta$ using a continuous set of measurements settings. Can this construction be extended to include additional settings?

An idea to explore more measurement settings would be to increase the space of nullifiers in which the squares are chosen. In~\cite{Bamps15}, the authors introduce a family of five linearly independent formal nullifiers of the partially entangled state $\ket{\phi_\theta}$ for a given choice of settings. These are used to find the SOS decomposition of previously known Bell inequalities including the tilted CHSH inequalities and more recently for the generalized tilted Bell inequalities~\cite{Li22}. In our case, we define nullifier operators independently of the measurement settings, thus leaving the choice of appropriate formal nullifiers to the SOS condition~\eqref{eq:SOScondition}.

In this section, we showcase how constructing squares in terms of additional nullifiers can lead to new Bell expressions suitable for more realizations. To do so we add a third nullifier to the two considered previously:
\begin{equation}
    \Hat N_2 = 1 - s_{2\theta} \Hat X_A \Hat X_B - c_{2\theta} \Hat Z_B
\end{equation}
Adding this third nullifier allows one to lift the conditions on Bob's angles to have symmetric measurements around the $Z$ axis, while Alice's measurements remains the same: we can set
\begin{equation}\label{eq:MoreSettingsPhiP}
    \left\{\begin{split}
        & \Hat M_1^{(1)} = \Hat Z_A, \quad \Hat M_1^{(1)} = \Hat X_A, \\
        & \Hat M_y^{(2)} = \cos(b_y)\Hat Z_B+\sin(b_y)\Hat X_B,
    \end{split} \right.
\end{equation}
where angles $b_1$ and $-b_2$ might be different. Note that up to relabelling of measurements incomes and/or outcomes, we can always assume that $b_1,b_2 \in (-\pi/2, \pi/2]$ and $b_1 < b_2$.

With those measurements, the corresponding formal nullifiers are given by:
\begin{subequations}
    \begin{align}
        N_0 = & A_1 - \frac{\sin(b_2)B_1 - \sin(b_1)B_2}{\sin(b_2-b_1)},\\
        N_1 = & A_2 - s_{2\theta} \frac{-\cos(b_2)B_1 + \cos(b_1)B_2}{\sin(b_2-b_1)}\\
        & - c_{2\theta} A_2 \frac{\sin(b_2)B_1 - \sin(b_1)B_2}{\sin(b_2-b_1)},\nonumber\\
        N_2 = &1 - s_{2\theta} A_2\frac{-\cos(b_2)B_1 + \cos(b_1)B_2}{\sin(b_2-b_1)} \\
        & - c_{2\theta} \frac{\sin(b_2)B_1 - \sin(b_1)B_2}{\sin(b_2-b_1)}.\nonumber
    \end{align}
\end{subequations}
We consider an SOS of the form $N_0^2 + (\lambda_1 N_1 + \lambda_2 N_2)^2$ for real parameters $\lambda_1$ and $\lambda_2$. When developing the squares we obtain terms proportional to the anticommutator $\{B_1,B_2\}$ and to $A_2\{B_1,B_2\}$ which contribute to $\Gamma$. The conditions $\Gamma=0$ thus leads to two equations:
\begin{subequations}
    \begin{align}
        \alpha  (\lambda_1^2+\lambda_2^2) + 2\beta \lambda_1 \lambda_2 &= 0,\\
        \beta (\lambda_1^2+\lambda_2^2) + 2\alpha\, \lambda_1 \lambda_2 &= - s_{b_1}s_{b_2},
    \end{align}
\end{subequations}
with
\begin{subequations}
    \begin{align}
        & \alpha = - \frac{1}{2} s_{4\theta}s_{b_1+b_2}\\
        & \beta = s_{2\theta}^2c_{b_1}c_{b_2}+c_{2\theta}^2s_{b_1}s_{b_2}.
    \end{align}
\end{subequations}

These two equations admit a solution when Bob's measurements lie in the squared region:
\begin{subequations}\label{eq:squareRegion}
\begin{align}
    b_1&\in (-2\theta,0)\\
    b_2&\in (0,2\theta).
\end{align}
\end{subequations}
The sum of square decompostion is then given by:
\begin{equation}
    N_0^2 + (\lambda_1 N_1 + \lambda_2 N_2)^2 = C(\theta,b_1,b_2) - S_{\theta,b_1,b_2}
\end{equation}
where: 
\begin{equation} \label{eq:ineqmoresettings}
    \scalebox{0.9}{$\begin{aligned}
        & \begin{aligned}
            S&_{\theta,b_1,b_2}  = 2 A_1\frac{s_{b_2}B_1 - s_{b_1}B_2}{s_{b_2-b_1}} \\
            & - 4\lambda_1\lambda_2\left[A_2 - s_{2\theta} \frac{-c_{b_2}B_1 +c_{b_1}B_2}{s_{b_2-b_1}} - c_{2\theta} A_2 \frac{s_{b_2}B_1 - s_{b_1}B_2}{s_{b_2-b_1}}\right] \\
            & + 2(\lambda_1^2 + \lambda_2^2)\left[ s_{2\theta} A_2 \frac{-c_{b_2}B_1 +c_{b_1}B_2}{s_{b_2-b_1}} + c_{2\theta} \frac{s_{b_2}B_1 - s_{b_1}B_2}{s_{b_2-b_1}} \right] ,
        \end{aligned}\\
        & C(\theta,b_1,b_2) = 2(1+\lambda_1^2+\lambda_2^2)
    \end{aligned}$}
\end{equation}
and 
\begin{subequations} \label{eq:lambdasexpressions}
    \begin{align}
        \lambda_1\lambda_2 &= -\frac{s_{b_1}s_{b_2} s_{b_1+b_2}s_{4\theta}}{(c_{2b_1} - c_{4\theta}) (c_{2b_2} - c_{4\theta})}, \\
        \lambda_1^2+\lambda_2^2 &= -\frac{4s_{b_1}^2s_{b_2}^2 (c_{2\theta}^2 + \cot(b_1)\cot(b_2)s_{2\theta}^2)}{(c_{2b_1} - c_{4\theta}) (c_{2b_2} - c_{4\theta})}. 
    \end{align}
\end{subequations}
The new candidate Bell expressions with associated quantum bound are given by:
\begin{equation}
    S_{\theta,b_1,b_2} \preceq C(\theta,b_1,b_2).
\end{equation}
These new expressions can also be used to self-test the partially entangled two qubit state $\ket{\phi_\theta}$ along with the measurements settings defined in \cref{eq:MoreSettingsPhiP}. The proof of the self-test can be found in Appendix \ref{sec:ProofPartial2}.

\begin{figure}
    \centering
    \includegraphics[width=0.5\textwidth]{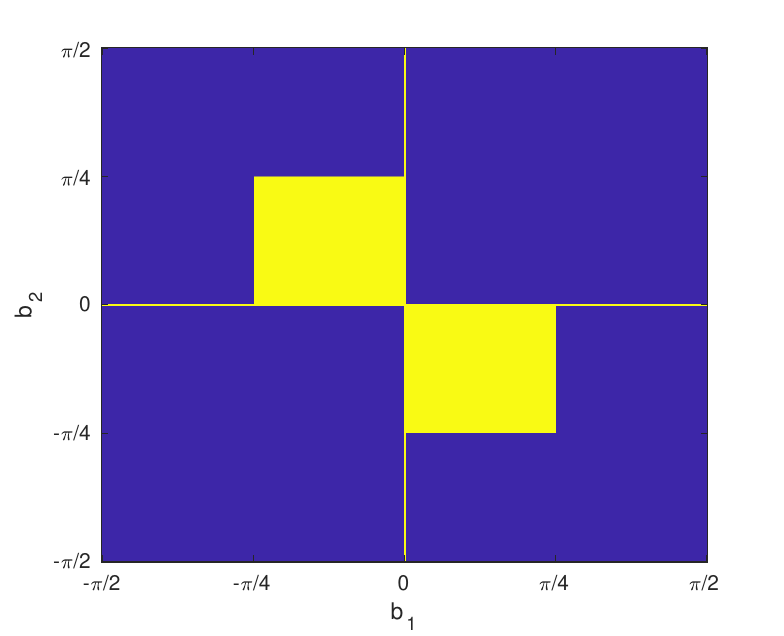}
    \caption{Result of the optimization \cref{eq:decomposability} for various choice of measurement angles for Bob. Points with $\Delta \leq 10^{-11}$ in yellow are on the boundary of the quantum set. Except when $b_1=0$ or $b_2=0$, in which case the behavior admits a probability equal to zero (both Alice and Bob measure in the $\hat Z$ direction), all points outside of the interval \cref{eq:squareRegion} belong to the blue region.}
    \label{fig:Decomposability}
\end{figure}

In the case where Alice's measurements are given by $\hat Z$ and $\hat X$, the settings given by \cref{eq:squareRegion} seem to be the only ones allowing to self-tested the partially-entangled state $\ket{\phi_\theta}$. Indeed, we verify numerically that any other choice of measurement settings for Bob results in behaviors that do not lie on the boundary of the NPA relaxation set at local level $\ell=1$. For this, we consider the following optimization
\begin{equation}\label{eq:decomposability}
\begin{split}
    \Delta = \underset{i}{\min}\ \underset{\bm P_1, \bm P_2}{\max}\ & \bm P_1^i - \bm P_2^i\\
    \text{s.t. } & \bm P_1^j=\bm P_2^j=\bm P^j,\ j\neq i\\
    & \bm P_1, \bm P_2 \in \text{NPA}_{\ell}
\end{split}
\end{equation}
Here $i$ runs over all components of the behavior $\bm P(ax|by)$ seen as a vector in $\mathbb{R}^8$ and $\text{NPA}_{\ell}$ stands for the $\ell^\text{th}$ level of the NPA hierarchy. The result of this optimization is zero iff the point is on the boundary of the NPA relaxation. \cref{fig:Decomposability} shows the result of this optimization as a function of the parameters $b_1$, $b_2$ in the case $\theta=\pi/8$. We see that all statistics outside the considered region \cref{eq:squareRegion} and its symmetric version for $B_1 \leftrightarrow B_2$ (i.e.~$b_1>b_2$) admit a decomposition.

\subsection{Multi-partite entangled states}\label{sec:multipartite}
In this section, we present a generalization of the expressions we found in the case of partially entangled two-qubit states whose maximum score is achieved by partially entangled GHZ states for an arbitrary number of parties $n$. We present only the results as the core of the reasoning follows the discussion of the subsection \ref{sec:partial1}.

The realizations that we aim to find a Bell expression for are given by the following combinations of the state and the measurements
\begin{equation} \label{eq:multisettings}
    \left\{\begin{split}
        & \ket{\psi} \sim \ket{\text{GHZ}_{n,\theta}} = c_\theta \ket{0...0} + s_\theta\ket{1...1} \\
        & \hat M_1^{(i)} = \hat Z^{(i)}, \quad \hat M_2^{(i)} = \hat X^{(i)}, \text{for }\ i < n\\
        & \hat M_y^{(n)} = \cos(b)\hat Z^{(n)}-(-1)^y\sin(b)\hat X^{(n)}.
    \end{split} \right.
\end{equation}
The family of nullifiers that we use contains $n$ operators given by
\begin{equation}
    \begin{split}
        &  \hat N_{0,i} = \ \hat Z^{(i)} - \hat Z^{(n)} \quad \forall i=1,\dots,n-1,\\
        & \hat N_1 = \prod_{i=1}^{n-1}\hat X^{(i)} - s_{2\theta}\hat X^{(n)} - c_{2\theta}\prod_{i=1}^{n-1}\hat X^{(i)}\cdot \hat Z^{(n)}
    \end{split}
\end{equation}
The sum of squares that we look at is of the form $\sum_i \Hat{N}_{0,i}^2 + \lambda^2 \Hat{N}_1^2$.  Under the condition $\frac{1}{\lambda^2}=s_{2\theta}^2\text{ct}^2(b) - c_{2\theta}^2$, just like in \cref{constrain}, we obtain the following SOS decomposition
\begin{equation} \label{eq:sosmoresettings}
    \sum_i N_{0,i}^2 + \lambda^2 N_1^2 = 2(1+\lambda^2)-S_{\theta,b}^n, 
\end{equation}
where
\begin{subequations}\label{eq:multibellineq}
    \begin{align}
        &N_{0,i} = Y_1^{(i)}\frac{Y_1^{(n)}+Y_2^{(n)}}{\cos(b)} \quad \forall \, i<n\\
        &N_1 = \prod_{i=1}^{n-1}Y_2^{(i)} - s_{2\theta}\frac{Y_1^{(2)}-Y_2^{(2)}}{\sin(b)} \\
        & \qquad \quad - c_{2\theta}\prod_{i=1}^{n-1}Y_2^{(i)} \cdot \frac{Y_1^{(n)}+Y_2^{(n)}}{\cos(b)} \nonumber\\
        & S_{\theta,b}^n = \frac{1}{n-1}\sum_{i=1}^{n-1}Y_1^{(i)}\frac{Y_1^{(n)}+Y_2^{(n)}}{\cos(b)} \\
        & \ + \lambda^2\left[s_{2\theta}\prod_{i=1}^{n-1}Y_2^{(i)}\cdot\frac{Y_1^{(2)}-Y_2^{(2)}}{\sin(b)} + c_{2\theta}\frac{Y_1^{(n)}+Y_2^{(n)}}{\cos(b)}\right].\nonumber
    \end{align}
\end{subequations}
The Tsirelson bound take the simple form
\begin{equation}
    S_{\theta,b}^n \preceq 2(1+\lambda^2).
\end{equation}  
In Appendix \ref{sec:ProofPartial} we prove that the saturation of these inequalities self-tests the state and the measurements in \cref{eq:multisettings}. 
Note that the case $n=2$ recovers exactly the result of section \ref{sec:partial1} and the inequality $S_{\theta,b} \preceq 2(1+\lambda^2)$.

\subsection{Maximally entangled state of two qutrits}\label{sec:qutrits}

Finally, we apply the SOS method to the case of the maximally entangled state of two qutrits
\begin{equation}
    \ket{\psi^3} = \frac{\ket{00}+\ket{11}+\ket{22}}{\sqrt{3}}.
\end{equation}
In general, it may be difficult to apply the method on a full parametrization of the measurement bases for large Hilbert space dimension. But the Bell expressions provided by the method are maximized by the desired state even when the measurement bases considered are not fully parametrized. We thus consider a family of measurements which generalizes the ones first used in~\cite{Collins02} to maximize the CGLMP expression, and then in~\cite{Salavrakos17,Sarkar21} to self-test the maximally entangled two qutrits state. Namely, for $x,y\in\{1,2\}$ we choose measurement bases of the form
\begin{subequations}
\begin{align}
    &\hat \Pi_{a|x} = U(a_x) F^\dagger \ketbra{a} F U(a_x)^\dagger \\
    &\hat \Pi_{b|y} = U(b_y) F^\dagger \ketbra{b} F U(b_y)^\dagger
\end{align}
\end{subequations}
These bases are related to the computational one by the Fourier transform 
\begin{equation}
    F = \frac{1}{\sqrt{3}} \sum_{k,l=0}^2 w^{kl} \ketbra{k}{l} \quad \text{with}\quad w=\exp(2i\pi/3)
\end{equation}
followed by a phase rotation with a real parameter $\theta$
\begin{equation}
    U(\theta) = \sum_{k=0}^2 w^{k\theta} \ketbra{k}.
\end{equation}

By analogy with the case of formal polynomials in \cref{eq: FT measuremnt opearators}, we define the unitary operators associated to each measurement by
\begin{subequations}
\begin{align}
    \hat M_x^{(1)} &= \sum_{a=0}^2 w^a \hat \Pi_{a|x}= \begin{pmatrix}
    0 & 0 & w^{-2a_x} \\
    w^{a_x} & 0 & 0 \\
    0 & w^{a_x} & 0 
    \end{pmatrix}
    \\
    \hat M_y^{(2)} &= \sum_{b=0}^2 w^b \hat \Pi_{b|y} = \begin{pmatrix}
    0 & 0 & w^{-2b_y} \\
    w^{b_y} & 0 & 0 \\
    0 & w^{b_y} & 0 
    \end{pmatrix}.
\end{align}
\end{subequations}
In the case of qutrits these operators verify $\left(\hat M_x^{(i)}\right)^\dagger = \left(\hat M_x^{(i)}\right)^2$, and any projector $\hat{\Pi}_{a|x}$ can be expressed as a linear combination of the identity operator, $\hat M^{(i)}_{x}$ and its adjoint $\left(\hat M_x^{(i)}\right)^\dagger $.

The first step of the SOS method is to find nullifiers for the target state. To do so we exploit the fact that for any unitary operator $\hat M$ the following identity holds: 
\begin{equation}
    \hat M\otimes \hat M^* \ket{\psi^3} = \ket{\psi^3}, 
\end{equation}
where $\hat M^*$ is the complex conjugate of $\hat M$. Thus any choice $\hat N = \id - \hat M\otimes \hat M^*$ with a unitary $\hat M$ defines a nullifier. Defining two nullifiers as
\begin{equation}
    \hat N_x = \id - \hat M_x^{(1)} \otimes \hat{\overbar{M}}_x^{(2)}, \ x\in\{1,2\},
\end{equation}
we obtain two operators $\hat{\overbar{M}}_x^{(2)}$ on Bob's Hilbert space that must verify
\begin{equation}
    \hat{\overbar{M}}_x^{(2)} = \left({{}\hat{M}_x^{(1)}}\right)^*
    \label{barcondition}
\end{equation}
when acting on Alice's.

The most general way to construct these operators from Bob's measurement operators is to take 
\begin{equation}
\begin{split}
    \hat{\overbar{M}}_x^{(2)} = &\ c_x\id + \mu_{1,x} \hat M_1^{(2)} + \mu_{2,x} \hat M_2^{(2)} \\
    &+\nu_{1,x} \left(\hat M_1^{(2)}\right)^2 +\nu_{2,x} \left(\hat M_2^{(2)}\right)^2.
\end{split}
\end{equation}
Now, together with equation~(\ref{barcondition}), this implies  that $c_x=\nu_{1,x}=\nu_{2,x}=0$ and 
\begin{equation}
    \left\{ \begin{split}
        \mu_{1,x} w^{b_1} + \mu_{2,x} w^{b_2} = w^{-a_x} \\
        \mu_{1,x} w^{-2b_1} + \mu_{2,x} w^{-2b_2} = w^{2a_x}.
    \end{split}\right.
\end{equation}
With simple algebra this system of equation solves to
\begin{equation}
    \left\{ \begin{split}
        \mu_{1,x} =\frac{w^{2a_x} - w^{-a_x - 3b_2}}{w^{-2b_1} - w^{b_1 - 3b_2}} \\
        \mu_{2,x} =\frac{w^{2a_x} - w^{-a_x - 3b_1}}{w^{-2b_2} - w^{b_2 - 3b_1}}.
    \end{split}\right.
    \label{muform}
\end{equation}

Now that we defined our two nullifiers we move to the formal polynomial formalism. We look at the formal polynomials associated to the nullifiers $N_x = \id - A_x \otimes \overbar{B_x}$ and the sum of squares $p N_1^\dagger N_1 + (1-p) N_2^\dagger N_2$ for an arbitrary $p\in(0,1)$ (the limit cases $p=0,1$ where only one nullifier appears would most likely be insufficient to grant self-testing). When developing the squares we get:  
\begin{equation}
    p N_1^\dagger N_1 + (1-p) N_2^\dagger N_2 = C - \Hat{S} + \Hat{\Gamma},
\end{equation}
where $C$=2, $S$ is a formal polynomial of local degree $1$ and $\Gamma$ is the leftover of higher local degree given here by
\begin{equation}
    \Gamma = \left(p\mu_{1,1}^\star \mu_{2,1} + (1-p)\mu_{1,2}^\star \mu_{2,2}\right)B_1^\dagger B_2 + \textit{h.c.}.
\end{equation}
Following the SOS method we look for the parameter regime where the leftover term  $\Gamma$ vanishes. This is the case when
\begin{equation}
    p\mu_{1,1}^\star \mu_{2,1} + (1-p)\mu_{1,2}^\star \mu_{2,2}=0.
\end{equation}
This can be written as a condition on the measurements parameters $a_x, b_y$:
\begin{equation}
\begin{split}
    & \begin{split}
        0= &\ p (w^{-2a_1} - w^{a_1 + 3b_2})(w^{2a_1} - w^{-a_1 - 3b_1}) \\ 
        &+ (1-p)(w^{-2a_2} - w^{a_2 + 3b_2})(w^{2a_2} - w^{-a_2 - 3b_1})
    \\
        \iff 0=&\ p (1 - w^{3a_1 + 3b_2})(1 - w^{-3a_1 - 3b_1}) \\ 
        &+ (1-p)(1 - w^{3a_2 + 3b_2})(1 - w^{-3a_2 - 3b_1}) 
    \\
        \iff 0=&\ p (1 - e^{2i\pi(a_1 + b_2)})(1 - e^{-2i\pi(a_1 + b_1)}) \\ 
        &+ (1-p)(1 - e^{2i\pi(a_2 + b_2)})(1 - e^{-2i\pi(a_2 + b_2)})
    \end{split} 
\end{split}
\end{equation}
leads to the condition
\begin{equation}\label{eq: p constraint qutrits}
    \begin{split}
       & p\sin(\pi(a_1 + b_2))\sin(\pi(a_1 + b_1)) \\
       & + (1-p)\sin(\pi(a_2 + b_2))\sin(\pi(a_2 + b_1)) = 0.
    \end{split}
\end{equation}

We thus obtain a family of Bell expressions with a maximal quantum violation given by: 
\begin{equation}
	S = p A_1 \overbar{B_1} + (1-p) A_2 \overbar{B_2} + \textit{h.c.} \preceq 2,
\end{equation}
where $\overbar{B_x} = \mu_{1,x} B_1 + \mu_{2,x} B_2$ and coefficient $\mu_{y,x}$ are given by equations~(\ref{muform}). Here, $a_x$, $b_y$ and $p$ are free parameters constrained only by \cref{eq: p constraint qutrits}. Without loss of generality we can set $a_1=0$ and $-b_1<-b_2$, and choose $a_2,-b_1,-b_2$ in $[0,\pi)$. \cref{eq: p constraint qutrits} then implies the alternating condition $-b_1<a_2<-b_2$ in analogy with the qubit case. Note that the average value of $S$ over any state is real as the Hermitian conjugate ($h.c.$) part ensures that $S$ is a Hermitian polynomial.

To further study these Bell expression candidates, we look at the second order of the variational method for small perturbation of the measurement parameters $a_x$, $b_y$. This analysis shows that the two negative eigenvalues of the Hessian matrix $\gamma$ are equal for measurements parameters $a_1 = 0, a_2 = 1/2, b_1 = 1/4, b_2 = 3/4$. Up to a local unitary transform, these are the parameters used in \cite{Salavrakos17}, which implies a form of optimality for this Bell expression within the considered family. As proven in~\cite{Salavrakos17,Sarkar21}, this inequality self-tests the maximally entangled state of two qutrits.

\section*{Conclusion}

In this work we considered the problem of constructing a Bell expression that is tailored to a generic target state in the sense that its maximal admissible value can be achieved by measuring the state. We presented a solution to this problem in the form of a systematic method applicable to arbitrary quantum states which uses a sum of square condition to define the Bell expression coefficients.

In principle, this method is \emph{asymptotically complete} in the sense that every Bell expression with the desired property can be obtained by starting from a complete enough set of formal nullifiers. When the degree of the nullifiers is bounded, the method provides \emph{sufficient} (but not necessary) conditions for a Bell expression to be maximally violated by the target state. Therefore, in all cases the method constructs Bell expressions with the guarantee that their maximal value is achieved by the desired state. The SOS method is thus complementary to the variational method, also presented here in details, which provides \emph{necessary} conditions for the maximal value of a Bell expression to be achieved by a target state.

In addition to providing a Bell expression with the desired property, a key feature of the SOS method is that it also grants, by construction, its sum of squares. This provides a first step towards the self-testing of the quantum realization. We confirmed this advantage with several examples.

Namely, using the SOS method we constructed a single family of Bell expressions able to self-test all partially-entangled n-partite states of the form $\ket{\text{GHZ}_{n,\theta}}=\cos\theta\ket{0...0}+\sin\theta\ket{1...1}$. We also recovered all the self-tests of the maximally entangled two-qubit state $\ket{\phi^+}=(\ket{00}+\ket{11})/\sqrt{2}$ with two binary measurements in terms of Bell inequalities, and proved that all self-tests in this scenario with degenerate measurements are non-exposed. We then used the method to derive a family of Bell expressions with 2 parameters self-testing the partially entangled state $\ket{\phi_\theta}=\cos\theta\ket{00}+\sin\theta\ket{11}$ when Alice performs $\hat Z$ and $\hat X$ measurements. In turn, this allowed us to demonstrate that the set of quantum correlations admits nonlocal angulous points. Finally, we demonstrated the generality of our method by constructing a family of Bell expressions for the maximally entangled two-qutrit state $\ket{\psi^3}=(\ket{00}+\ket{11}+\ket{22})/\sqrt{3}$, inferring a form of optimality for the Bell expression introduced in~\cite{Salavrakos17}.

The constraints imposed by the SOS method apply to both the state and measurements parameters. Therefore, the method can be used to construct inequalities tailored to target measurements as well, such as families of Bell expressions for fixed settings and varying states~\cite{Li22}. It would be interesting to further investigate the relevance of the SOS method to self-testing of measurements.

Another open question would be to clarify when Bell expressions obtained by the SOS method exhibit the self-testing property. Whereas SOS decompositions provide a key ingredient to self-testing, complete self-testing proofs can still require substantial work~\cite{Sarkar21,Sarkar22}. Finding conditions under which the SOS method allows for self-testing could lead to an asymptotically complete method for self-testing target states. 

Finally, it would be interesting to better understand the resistance to noise of the obtained self-tests. Given the families of Bell expressions able to self-test the partially entangled state of two qubits that we discovered with the SOS method, it is natural to ask which of these is most robust to noise. In fact, as we showed that several Bell expressions can sometimes self-test the same point of correlations, it would be interesting to answer this question even for a fixed set of measurement settings.

\acknowledgements
We acknowledge funding by Commissariat à l’Energie Atomique et aux Energies Alternatives (CEA), the European Union’s Horizon 2020 research and innovation program European High-Performance Computing Joint Undertaking under grant agreement No 101018180 (HPCQS) and a French national quantum initiative managed by Agence Nationale de la Recherche in the framework of France 2030 with the reference ANR- 22-PETQ-0007. P.S. acknowledges funding from the Swiss National Science Foundation (project 192244).

\bibliographystyle{quantum}
\bibliography{bibliographie}

\providecommand{\noopsort}[1]{}\providecommand{\singleletter}[1]{#1}%
\begin{thebibliography}{10}

\bibitem{Brunner14}
Nicolas Brunner, Daniel Cavalcanti, Stefano Pironio, Valerio Scarani, and Stephanie Wehner.
\newblock ``Bell nonlocality''.
\newblock \href{https://dx.doi.org/10.1103/RevModPhys.86.419}{Rev. Mod. Phys. {\bf 86}, 419--478}~(2014).

\bibitem{Brassard05}
Gilles Brassard, Anne Broadbent, and Alain Tapp.
\newblock ``{Quantum Pseudo-Telepathy}''.
\newblock \href{https://dx.doi.org/10.1007/s10701-005-7353-4}{Found. Phys. {\bf 35}, 1877--1907}~(2005).

\bibitem{Werner89}
Reinhard~F. Werner.
\newblock ``{Quantum states with Einstein-Podolsky-Rosen correlations admitting a hidden-variable model}''.
\newblock \href{https://dx.doi.org/10.1103/PhysRevA.40.4277}{Phys. Rev. A {\bf 40}, 4277--4281}~(1989).

\bibitem{Bancal11}
Jean-Daniel Bancal, Nicolas Gisin, Yeong-Cherng Liang, and Stefano Pironio.
\newblock ``{Device-Independent Witnesses of Genuine Multipartite Entanglement}''.
\newblock \href{https://dx.doi.org/10.1103/PhysRevLett.106.250404}{Phys. Rev. Lett. {\bf 106}, 250404}~(2011).

\bibitem{Colbeck06}
Roger Colbeck.
\newblock ``{Quantum and relativistic protocols for secure multi-party computation}''.
\newblock \href{https://dx.doi.org/10.48550/arXiv.0911.3814}{PhD thesis}.
\newblock University of Cambridge.
\newblock ~(2006).

\bibitem{Colbeck11}
Roger Colbeck and Adrian Kent.
\newblock ``{Private randomness expansion with untrusted devices}''.
\newblock \href{https://dx.doi.org/10.1088/1751-8113/44/9/095305}{Journal of Physics A: Mathematical and Theoretical {\bf 44}, 095305}~(2011).

\bibitem{Pironio10}
S.~Pironio, A.~Acín, S.~Massar, A.~Boyer de~la Giroday, D.~N. Matsukevich, P.~Maunz, S.~Olmschenk, D.~Hayes, L.~Luo, T.~A. Manning, and C.~Monroe.
\newblock ``{Random numbers certified by Bell’s theorem}''.
\newblock \href{https://dx.doi.org/10.1038/nature09008}{nature {\bf 464}, 1021--1024}~(2010).

\bibitem{Mayers04}
Dominic Mayers and Andrew Yao.
\newblock ``Self testing quantum apparatus''.
\newblock \href{https://dx.doi.org/10.26421/QIC4.4-3}{Quantum information \& computation~{\bf 4}}~(2003).

\bibitem{Acin07}
Antonio Ac\'{\i}n, Nicolas Brunner, Nicolas Gisin, Serge Massar, Stefano Pironio, and Valerio Scarani.
\newblock ``{Device-Independent Security of Quantum Cryptography against Collective Attacks}''.
\newblock \href{https://dx.doi.org/10.1103/PhysRevLett.98.230501}{Phys. Rev. Lett. {\bf 98}, 230501}~(2007).

\bibitem{Ekert91}
Artur~K. Ekert.
\newblock ``{Quantum cryptography based on Bell's theorem}''.
\newblock \href{https://dx.doi.org/10.1103/PhysRevLett.67.661}{Phys. Rev. Lett. {\bf 67}, 661--663}~(1991).

\bibitem{Nadlinger22}
D.~P. Nadlinger, P.~Drmota, B.~C. Nichol, G.~Araneda, D.~Main, R.~Srinivas, D.~M. Lucas, C.~J. Ballance, K.~Ivanov, E.~Y.-Z. Tan, P.~Sekatski, R.~L. Urbanke, R.~Renner, N.~Sangouard, and J.-D. Bancal.
\newblock ``{Experimental quantum key distribution certified by Bell's theorem}''.
\newblock \href{https://dx.doi.org/10.1038/s41586-022-04941-5}{nature {\bf 607}, 682--686}~(2022).

\bibitem{Zhang22}
Wei Zhang, Tim van Leent, Kai Redeker, Robert Garthoff, René Schwonnek, Florian Fertig, Sebastian Eppelt, Wenjamin Rosenfeld, Valerio Scarani, Charles C.-W. Lim, and Harald Weinfurter.
\newblock ``{A device-independent quantum key distribution system for distant users}''.
\newblock \href{https://dx.doi.org/10.1038/s41586-022-04891-y}{nature {\bf 607}, 687--691}~(2022).

\bibitem{Liu22}
Wen-Zhao Liu, Yu-Zhe Zhang, Yi-Zheng Zhen, Ming-Han Li, Yang Liu, Jingyun Fan, Feihu Xu, Qiang Zhang, and Jian-Wei Pan.
\newblock ``{Toward a Photonic Demonstration of Device-Independent Quantum Key Distribution}''.
\newblock \href{https://dx.doi.org/10.1103/PhysRevLett.129.050502}{Phys. Rev. Lett. {\bf 129}, 050502}~(2022).

\bibitem{Pitowsky89}
Itamar Pitowsky.
\newblock ``{Classical Correlation Polytopes and Propositional Logic}''.
\newblock In {Quantum Probability \textemdash{} {{Quantum}} Logic}.
\newblock \href{https://dx.doi.org/10.1007/BFb0021188}{Pages 11--51}.
\newblock {Springer Berlin Heidelberg}, {Berlin, Heidelberg}~(1989).

\bibitem{Clauser69}
J.~F. Clauser, M.~A. Horne, A.~Shimony, and R.~A. Holt.
\newblock ``{Proposed Experiment to Test Local Hidden-Variable Theories}''.
\newblock \href{https://dx.doi.org/10.1103/PhysRevLett.23.880}{Phys. Rev. Lett. {\bf 23}, 880}~(1969).

\bibitem{Kaszlikowski02}
Dagomir Kaszlikowski, L.~C. Kwek, Jing-Ling Chen, Marek \ifmmode~\dot{Z}\else \.{Z}\fi{}ukowski, and C.~H. Oh.
\newblock ``{Clauser-Horne inequality for three-state systems}''.
\newblock \href{https://dx.doi.org/10.1103/PhysRevA.65.032118}{Phys. Rev. A {\bf 65}, 032118}~(2002).

\bibitem{Collins02}
Daniel Collins, Nicolas Gisin, Noah Linden, Serge Massar, and Sandu Popescu.
\newblock ``{Bell Inequalities for Arbitrarily High-Dimensional Systems}''.
\newblock \href{https://dx.doi.org/10.1103/PhysRevLett.88.040404}{Phys. Rev. Lett. {\bf 88}, 040404}~(2002).

\bibitem{Acin02}
A.~Ac\'{\i}n, T.~Durt, N.~Gisin, and J.~I. Latorre.
\newblock ``{Quantum nonlocality in two three-level systems}''.
\newblock \href{https://dx.doi.org/10.1103/PhysRevA.65.052325}{Phys. Rev. A {\bf 65}, 052325}~(2002).

\bibitem{Navascues07}
Miguel Navascu{\'e}s, Stefano Pironio, and Antonio Ac{\'i}n.
\newblock ``{Bounding the {{Set}} of {{Quantum Correlations}}}''.
\newblock \href{https://dx.doi.org/10.1103/PhysRevLett.98.010401}{Physical Review Letters {\bf 98}, 010401}~(2007).

\bibitem{Ioannou22}
Marie Ioannou and Denis Rosset.
\newblock ``{Noncommutative polynomial optimization under symmetry}''~(2022).
\newblock  \href{http://arxiv.org/abs/2112.10803}{arXiv:2112.10803}.

\bibitem{Ji08}
Se-Wan Ji, Jinhyoung Lee, James Lim, Koji Nagata, and Hai-Woong Lee.
\newblock ``{Multisetting {{Bell}} Inequality for Qudits}''.
\newblock \href{https://dx.doi.org/10.1103/PhysRevA.78.052103}{Physical Review A {\bf 78}, 052103}~(2008).

\bibitem{Liang09}
Yeong-Cherng Liang, Chu-Wee Lim, and Dong-Ling Deng.
\newblock ``Reexamination of a multisetting {{Bell}} inequality for qudits''.
\newblock \href{https://dx.doi.org/10.1103/PhysRevA.80.052116}{Physical Review A {\bf 80}, 052116}~(2009).

\bibitem{Lim10}
James Lim, Junghee Ryu, Seokwon Yoo, Changhyoup Lee, Jeongho Bang, and Jinhyoung Lee.
\newblock ``{Genuinely High-Dimensional Nonlocality Optimized by Complementary Measurements}''.
\newblock \href{https://dx.doi.org/10.1088/1367-2630/12/10/103012}{New Journal of Physics {\bf 12}, 103012}~(2010).

\bibitem{Salavrakos17}
Alexia Salavrakos, Remigiusz Augusiak, Jordi Tura, Peter Wittek, Antonio Ac\'{\i}n, and Stefano Pironio.
\newblock ``{Bell Inequalities Tailored to Maximally Entangled States}''.
\newblock \href{https://dx.doi.org/10.1103/PhysRevLett.119.040402}{Phys. Rev. Lett. {\bf 119}, 040402}~(2017).

\bibitem{Acin12}
Antonio Ac{\'i}n, Serge Massar, and Stefano Pironio.
\newblock ``{Randomness versus {{Nonlocality}} and {{Entanglement}}}''.
\newblock \href{https://dx.doi.org/10.1103/PhysRevLett.108.100402}{Physical Review Letters {\bf 108}, 100402}~(2012).

\bibitem{Pironio03}
Stefano Pironio.
\newblock ``{Violations of {{Bell}} Inequalities as Lower Bounds on the Communication Cost of Nonlocal Correlations}''.
\newblock \href{https://dx.doi.org/10.1103/PhysRevA.68.062102}{Physical Review A {\bf 68}, 062102}~(2003).

\bibitem{NietoSilleras14}
O.~{Nieto-Silleras}, S.~Pironio, and J.~Silman.
\newblock ``{Using Complete Measurement Statistics for Optimal Device-Independent Randomness Evaluation}''.
\newblock \href{https://dx.doi.org/10.1088/1367-2630/16/1/013035}{New Journal of Physics {\bf 16}, 013035}~(2014).

\bibitem{Bancal14}
Jean-Daniel Bancal, Lana Sheridan, and Valerio Scarani.
\newblock ``{More Randomness from the Same Data}''.
\newblock \href{https://dx.doi.org/10.1088/1367-2630/16/3/033011}{New Journal of Physics {\bf 16}, 033011}~(2014).

\bibitem{Brown23}
Peter Brown, Hamza Fawzi, and Omar Fawzi.
\newblock ``Device-independent lower bounds on the conditional von neumann entropy''~(2023).
\newblock  \href{http://arxiv.org/abs/2106.13692}{arXiv:2106.13692}.

\bibitem{Navascues08}
Miguel Navascués, Stefano Pironio, and Antonio Acín.
\newblock ``{A convergent hierarchy of semidefinite programs characterizing the set of quantum correlations}''.
\newblock \href{https://dx.doi.org/10.1088/1367-2630/10/7/073013}{New Journal of Physics {\bf 10}, 073013}~(2008).

\bibitem{Popescu92}
Sandu Popescu and Daniel Rohrlich.
\newblock ``{Which States Violate {{Bell}}'s Inequality Maximally?}''.
\newblock \href{https://dx.doi.org/10.1016/0375-9601(92)90819-8}{Physics Letters A {\bf 169}, 411--414}~(1992).

\bibitem{McKague12}
M~McKague, T~H Yang, and V~Scarani.
\newblock ``Robust self-testing of the singlet''.
\newblock \href{https://dx.doi.org/10.1088/1751-8113/45/45/455304}{Journal of Physics A: Mathematical and Theoretical {\bf 45}, 455304}~(2012).

\bibitem{Supic20}
Ivan {\v{S}}upi{\'{c}} and Joseph Bowles.
\newblock ``{Self-testing of quantum systems: a review}''.
\newblock \href{https://dx.doi.org/10.22331/q-2020-09-30-337}{{Quantum} {\bf 4}, 337}~(2020).

\bibitem{Bamps15}
C\'edric Bamps and Stefano Pironio.
\newblock ``{Sum-of-squares decompositions for a family of Clauser-Horne-Shimony-Holt-like inequalities and their application to self-testing}''.
\newblock \href{https://dx.doi.org/10.1103/PhysRevA.91.052111}{Phys. Rev. A {\bf 91}, 052111}~(2015).

\bibitem{Sarkar21}
ShubhayanAU Sarkar, Debashis Saha, Jędrzej Kaniewski, and Remigiusz Augusiak.
\newblock ``{Self-testing quantum systems of arbitrary local dimension with minimal number of measurements}''.
\newblock \href{https://dx.doi.org/10.1038/s41534-021-00490-3}{npj Quantum Information {\bf 7}, 151}~(2021).

\bibitem{Coladangelo17}
Andrea Coladangelo, Koon~Tong Goh, and Valerio Scarani.
\newblock ``{All pure bipartite entangled states can be self-tested}''.
\newblock \href{https://dx.doi.org/10.1038/ncomms15485}{nature comm {\bf 8}, 15485}~(2017).

\bibitem{Wu14}
Xingyao Wu, Yu~Cai, Tzyh~Haur Yang, Huy~Nguyen Le, Jean-Daniel Bancal, and Valerio Scarani.
\newblock ``{Robust self-testing of the three-qubit $W$ state}''.
\newblock \href{https://dx.doi.org/10.1103/PhysRevA.90.042339}{Phys. Rev. A {\bf 90}, 042339}~(2014).

\bibitem{Pal14}
K\'aroly~F. P\'al, Tam\'as V\'ertesi, and Miguel Navascu\'es.
\newblock ``{Device-independent tomography of multipartite quantum states}''.
\newblock \href{https://dx.doi.org/10.1103/PhysRevA.90.042340}{Phys. Rev. A {\bf 90}, 042340}~(2014).

\bibitem{McKague14}
Matthew McKague.
\newblock ``Self-{{Testing Graph States}}''.
\newblock In Dave Bacon, Miguel {Martin-Delgado}, and Martin Roetteler, editors, Theory of {{Quantum Computation}}, {{Communication}}, and {{Cryptography}}.
\newblock \href{https://dx.doi.org/10.1007/978-3-642-54429-3_7}{Pages 104--120}.
\newblock Berlin, Heidelberg~(2014). Springer.

\bibitem{Supic18}
I~{\v{S}}upi{\'{c}}, A~Coladangelo, R~Augusiak, and A~Ac{\'{\i}}n.
\newblock ``{Self-testing multipartite entangled states through projections onto two systems}''.
\newblock \href{https://dx.doi.org/10.1088/1367-2630/aad89b}{New Journal of Physics {\bf 20}, 083041}~(2018).

\bibitem{Baccari19}
F.~Baccari, R.~Augusiak, I.~\ifmmode \check{S}\else \v{S}\fi{}upi\ifmmode~\acute{c}\else \'{c}\fi{}, J.~Tura, and A.~Ac\'{\i}n.
\newblock ``{Scalable Bell Inequalities for Qubit Graph States and Robust Self-Testing}''.
\newblock \href{https://dx.doi.org/10.1103/PhysRevLett.124.020402}{Phys. Rev. Lett. {\bf 124}, 020402}~(2020).

\bibitem{Supic22}
Ivan Šupić, Joseph Bowles, Marc-Olivier Renou, Antonio Acín, and Matty~J. Hoban.
\newblock ``Quantum networks self-test all entangled states''.
\newblock \href{https://dx.doi.org/10.1038/s41567-023-01945-4}{Nature Physics {\bf 19}, 670–675}~(2023).

\bibitem{Sarkar22}
Shubhayan Sarkar and Remigiusz Augusiak.
\newblock ``{Self-testing of multipartite Greenberger-Horne-Zeilinger states of arbitrary local dimension with arbitrary number of measurements per party}''.
\newblock \href{https://dx.doi.org/10.1103/PhysRevA.105.032416}{Phys. Rev. A {\bf 105}, 032416}~(2022).

\bibitem{Yang14}
Tzyh~Haur Yang, Tam{\'a}s V{\'e}rtesi, Jean-Daniel Bancal, Valerio Scarani, and Miguel Navascu{\'e}s.
\newblock ``{Robust and {{Versatile Black-Box Certification}} of {{Quantum Devices}}}''.
\newblock \href{https://dx.doi.org/10.1103/PhysRevLett.113.040401}{Physical Review Letters {\bf 113}, 040401}~(2014).

\bibitem{Rai22}
Ashutosh Rai, Matej Pivoluska, Souradeep Sasmal, Manik Banik, Sibasish Ghosh, and Martin Plesch.
\newblock ``{Self-Testing Quantum States via Nonmaximal Violation in {{Hardy}}'s Test of Nonlocality}''.
\newblock \href{https://dx.doi.org/10.1103/PhysRevA.105.052227}{Physical Review A {\bf 105}, 052227}~(2022).

\bibitem{Goh18}
Koon~Tong Goh, J\ifmmode \mbox{\k{e}}\else~\k{e}\fi{}drzej Kaniewski, Elie Wolfe, Tam\'as V\'ertesi, Xingyao Wu, Yu~Cai, Yeong-Cherng Liang, and Valerio Scarani.
\newblock ``{Geometry of the set of quantum correlations}''.
\newblock \href{https://dx.doi.org/10.1103/PhysRevA.97.022104}{Phys. Rev. A {\bf 97}, 022104}~(2018).

\bibitem{Chen23}
Kai-Siang Chen, Gelo Noel~M. Tabia, Chellasamy Jebarathinam, Shiladitya Mal, Jun-Yi Wu, and Yeong-Cherng Liang.
\newblock ``Quantum correlations on the no-signaling boundary: self-testing and more''.
\newblock \href{https://dx.doi.org/10.22331/q-2023-07-11-1054}{{Quantum} {\bf 7}, 1054}~(2023).

\bibitem{Coladangelo18}
Andrea Coladangelo.
\newblock ``{Generalization of the {{Clauser-Horne-Shimony-Holt}} Inequality Self-Testing Maximally Entangled States of Any Local Dimension}''.
\newblock \href{https://dx.doi.org/10.1103/PhysRevA.98.052115}{Physical Review A {\bf 98}, 052115}~(2018).

\bibitem{Sainz16}
Ana~Bel{\'e}n Sainz, Yelena Guryanova, Will McCutcheon, and Paul Skrzypczyk.
\newblock ``{Adjusting Inequalities for Detection-Loophole-Free Steering Experiments}''.
\newblock \href{https://dx.doi.org/10.1103/PhysRevA.94.032122}{Physical Review A {\bf 94}, 032122}~(2016).

\bibitem{Zwerger19}
M.~Zwerger, W.~D\"ur, J.-D. Bancal, and P.~Sekatski.
\newblock ``{Device-Independent Detection of Genuine Multipartite Entanglement for All Pure States}''.
\newblock \href{https://dx.doi.org/10.1103/PhysRevLett.122.060502}{Phys. Rev. Lett. {\bf 122}, 060502}~(2019).

\bibitem{Supic16}
I.~{\v S}upi{\'c}, R.~Augusiak, A.~Salavrakos, and A.~Ac{\'i}n.
\newblock ``Self-testing protocols based on the chained {{Bell}} inequalities''.
\newblock \href{https://dx.doi.org/10.1088/1367-2630/18/3/035013}{New Journal of Physics {\bf 18}, 035013}~(2016).

\bibitem{Doherty08}
Andrew~C. Doherty, Yeong-Cherng Liang, Ben Toner, and Stephanie Wehner.
\newblock ``The quantum moment problem and bounds on entangled multi-prover games''.
\newblock In Proceedings of the 2008 IEEE 23rd Annual Conference on Computational Complexity.
\newblock \href{https://dx.doi.org/10.1109/CCC.2008.26}{Page 199–210}.
\newblock CCC '08USA~(2008). IEEE Computer Society.

\bibitem{Augusiak19}
R~Augusiak, A~Salavrakos, J~Tura, and A~Acín.
\newblock ``{Bell inequalities tailored to the Greenberger–Horne–Zeilinger states of arbitrary local dimension}''.
\newblock \href{https://dx.doi.org/10.1088/1367-2630/ab4d9f}{New Journal of Physics {\bf 21}, 113001}~(2019).

\bibitem{Kaniewski19}
J{\k{e}}drzej Kaniewski, Ivan {\v{S}}upi{\'{c}}, Jordi Tura, Flavio Baccari, Alexia Salavrakos, and Remigiusz Augusiak.
\newblock ``{Maximal nonlocality from maximal entanglement and mutually unbiased bases, and self-testing of two-qutrit quantum systems}''.
\newblock \href{https://dx.doi.org/10.22331/q-2019-10-24-198}{{Quantum} {\bf 3}, 198}~(2019).

\bibitem{Santos23}
Rafael Santos, Debashis Saha, Flavio Baccari, and Remigiusz Augusiak.
\newblock ``Scalable {{Bell}} inequalities for graph states of arbitrary prime local dimension and self-testing''.
\newblock \href{https://dx.doi.org/10.1088/1367-2630/acd9e3}{New Journal of Physics {\bf 25}, 063018}~(2023).

\bibitem{Sekatski18}
Pavel Sekatski, Jean-Daniel Bancal, Sebastian Wagner, and Nicolas Sangouard.
\newblock ``Certifying the {{Building Blocks}} of {{Quantum Computers}} from {{Bell}}'s {{Theorem}}''.
\newblock \href{https://dx.doi.org/10.1103/PhysRevLett.121.180505}{Physical Review Letters {\bf 121}, 180505}~(2018).

\bibitem{Rosset14}
Denis Rosset, Jean-Daniel Bancal, and Nicolas Gisin.
\newblock ``{Classifying 50 years of Bell inequalities}''.
\newblock \href{https://dx.doi.org/10.1088/1751-8113/47/42/424022}{Journal of Physics A: Mathematical and Theoretical {\bf 47}, 424022}~(2014).

\bibitem{Scarani01}
Valerio Scarani and Nicolas Gisin.
\newblock ``{Spectral decomposition of {Bell}'s operators for qubits}''.
\newblock \href{https://dx.doi.org/10.1088/0305-4470/34/30/314}{Journal of Physics A: Mathematical and General {\bf 34}, 6043--6053}~(2001).

\bibitem{Guehne09}
Otfried Gühne and Géza Tóth.
\newblock ``Entanglement detection''.
\newblock \href{https://dx.doi.org/https://doi.org/10.1016/j.physrep.2009.02.004}{Physics Reports {\bf 474}, 1--75}~(2009).

\bibitem{Wagner20}
Sebastian Wagner, Jean-Daniel Bancal, Nicolas Sangouard, and Pavel Sekatski.
\newblock ``{Device-Independent Characterization of Quantum Instruments}''.
\newblock \href{https://dx.doi.org/10.22331/q-2020-03-19-243}{Quantum {\bf 4}, 243}~(2020).

\bibitem{eigenvalueperturbation}
``Eigenvalue perturbation, {Wikipedia, The Free Encyclopedia, Wikimedia Foundation}, 12 {July}''.
\newblock  url:~\url{https://en.wikipedia.org/wiki/Eigenvalue_perturbation}.

\bibitem{Pironio10b}
S.~Pironio, M.~Navascu\'{e}s, and A.~Ac\'{\i}n.
\newblock ``Convergent relaxations of polynomial optimization problems with noncommuting variables''.
\newblock \href{https://dx.doi.org/10.1137/090760155}{SIAM Journal on Optimization {\bf 20}, 2157--2180}~(2010).

\bibitem{Moroder13}
Tobias Moroder, Jean-Daniel Bancal, Yeong-Cherng Liang, Martin Hofmann, and Otfried G\"uhne.
\newblock ``Device-independent entanglement quantification and related applications''.
\newblock \href{https://dx.doi.org/10.1103/PhysRevLett.111.030501}{Phys. Rev. Lett. {\bf 111}, 030501}~(2013).

\bibitem{Reichardt13}
Ben~W. Reichardt, Falk Unger, and Umesh Vazirani.
\newblock ``Classical command of quantum systems''.
\newblock \href{https://dx.doi.org/10.1038/nature12035}{Nature {\bf 496}, 456--460}~(2013).

\bibitem{Wang16}
Yukun Wang, Xingyao Wu, and Valerio Scarani.
\newblock ``{All the self-testings of the singlet for two binary measurements}''.
\newblock \href{https://dx.doi.org/10.1088/1367-2630/18/2/025021}{New Journal of Physics {\bf 18}, 025021}~(2016).

\bibitem{Coopmans17}
Tim Coopmans, J\ifmmode \mbox{\k{e}}\else~\k{e}\fi{}drzej Kaniewski, and Christian Schaffner.
\newblock ``{Robust self-testing of two-qubit states}''.
\newblock \href{https://dx.doi.org/10.1103/PhysRevA.99.052123}{Phys. Rev. A {\bf 99}, 052123}~(2019).

\bibitem{Li22}
Xinhui Li, Yukun Wang, Yunguang Han, and Shi-Ning Zhu.
\newblock ``Self-testing of different entanglement resources via fixed measurement settings''.
\newblock \href{https://dx.doi.org/10.1103/physreva.106.052418}{Physical Review A{\bf 106}}~(2022).

\bibitem{Tsirelson93}
B.~S. Tsirelson.
\newblock ``{Some results and problems on quantum Bell-type inequalities}''.
\newblock Hadronic Journal Supplement {\bf 8}, 329--345~(1993).
\newblock  url:~\url{https://ci.nii.ac.jp/naid/10026857475/en/}.

\bibitem{Le23}
Thinh~P. Le, Chiara Meroni, Bernd Sturmfels, Reinhard~F. Werner, and Timo Ziegler.
\newblock ``Quantum {C}orrelations in the {M}inimal {S}cenario''.
\newblock \href{https://dx.doi.org/10.22331/q-2023-03-16-947}{{Quantum} {\bf 7}, 947}~(2023).

\end{thebibliography}

\appendix
\onecolumngrid

\section*{Appendices}

\section{Correlations from $\phi^+$ with one measurement in common are non-exposed} \label{sec:Limitpoints}
We address here the nature of the points for which Alice and Bob use one common measurement. In \cite{Goh18}, one of those points was studied to prove that there exist non exposed extremal points on the boundary of the quantum set of correlations. While showing it numerically on a plot, the authors gave a proof using the analytical characterization of quantum full-correlation by Tsirelson, Landau and Masanes \cite{Tsirelson93} stating that a behavior with uniform marginals $\langle A_x\rangle=\langle B_y\rangle=0$ belongs to the quantum set if and only if 
\begin{equation}
1+\prod_{x y}\left\langle A_{x} B_{y}\right\rangle+\prod_{x y} \sqrt{1-\left\langle A_{x} B_{y}\right\rangle^{2}}-\frac{1}{2} \sum_{x y}\left\langle A_{x} B_{y}\right\rangle^{2} \geqslant 0.
\label{tsi}
\end{equation}
We now use this characterization to prove that all correlations obtained with the $\ket{\phi^+}$ state and Alice and Bob having one common measurement is a non-exposed boundary point of the quantum set.

By non-exposed, we mean that there exist no tangent hyperplane to the quantum set of correlation that contains this point and only this point. Any such correlation can be written, up to relabelling measurements and/or outcomes, as
\begin{equation}
    \mathcal{P}_{x,y} = \begin{array}{c|c|c}
         & 0 & 0 \\
         \hline
         0 & 1 & \cos(x+y) \\
         \hline
         0 & \cos(x) & \cos(y)
    \end{array}
\end{equation}
where $0<x,y,x+y<\pi$. This point is on the boundary as it maximizes the inequality $\langle A_0B_0 \rangle \leq 1$. Moreover, it saturates inequality \cref{tsi}.

Let us introduce the classical point: 
\begin{equation}
    \mathcal{P}_{C} = \begin{array}{c|c|c}
         & 0 & 0 \\
         \hline
         0 & 1 & 0 \\
         \hline
         0 & 0 & 0
    \end{array}
\end{equation}
By convexity of the set of quantum correlations, any behavior on the segment $[\mathcal{P}_C,\mathcal{P}_{x,y}]$ is achievable and form a flat border of the set. What we aim to prove now is that any tangent hyperplane of the QSC containing $\mathcal{P}_{x,y}$, also contains this segment. This would prove that the point is non-exposed.

This segment can be parametrized as such: 
\begin{equation}
    \mathcal{P}_{x,y}(0\leq\alpha\leq1) = \begin{array}{c|c|c}
         & 0 & 0 \\
         \hline
         0 & 1 & \alpha\cos(x+y) \\
         \hline
         0 & \alpha\cos(x) & \alpha\cos(y).
    \end{array}
\end{equation}
We now show that there exist quantum correlation points that approach this line even for $\alpha \geq 1$. This implies that the slope of the border in the direction $\mathcal{P}_{x,y}-\mathcal{P}_{C}$ at the point $\mathcal{P}_{x,y}$ is defined and of value 0.

\begin{figure}
\begin{center}
    \includegraphics[width=0.4\textwidth]{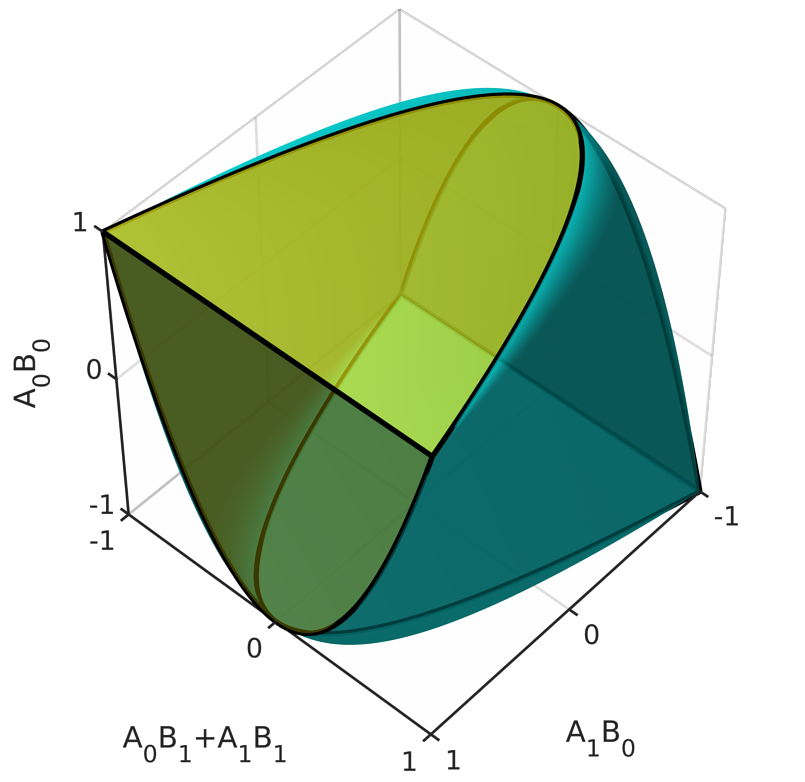}
\end{center}
\caption{\label{3dplot}Three dimensional slice of the quantum set of correlation (see also~\cite{Le23}). The points $\mathcal{P}_{x,x}$ for $x$ varying from $0$ to $\pi$ are on a black parabola. The light green planes are common boundary planes of the quantum set and the non-signaling set. The quantum set drops off smoothly from this plane at every point of the parabolas, except for their very summits, corresponding to a classical points.}
\end{figure}

To do so, let's consider a point written as: 
\begin{equation}
    \mathcal{P}_{x,y}(\alpha,\beta(\alpha)) = \begin{array}{c|c|c}
         & 0 & 0 \\
         \hline
         0 & \beta(\alpha) & \alpha\cos(x+y) \\
         \hline
         0 & \alpha\cos(x) & \alpha\cos(y)
    \end{array}
\end{equation}
for $\alpha\geq1$. We want to prove that when $\alpha=1+\text{d}\alpha$, there exist a set a quantum points for which $\beta(\alpha)=1-0\cdot\text{d}\alpha-\beta_2\text{d}\alpha^2$. The function of the correlation on the left hand side of \cref{tsi} is, at first order in $\text{d}\alpha$: 
\begin{equation}
    (g(x,y) + \sqrt{2\beta_2})\text{d}\alpha + o(\text{d}\alpha^2),
\end{equation}
where $g(x,y)$ is some function of $x$ and $y$. Thus for a sufficiently large value of $\beta_2$, the left hand side term of \cref{tsi} is positive and the points $\mathcal{P}_{x,y}(1+\text{d}\alpha,1-\beta_2\text{d}\alpha^2)$ belongs -- for small $\text{d}\alpha$ -- to the quantum set. This proves our result. The non-exposed property of those points in the case $y=x$ is showcased in \cref{3dplot}.

\section{Self-testing for the candidate expressions}

Here we prove self-testing for most of the Bell expressions constructed in the main text. Before considering specific cases, we describe how the SOS method provides tools for self-testing.

\subsection{How to prove the self-test: analytical relations}\label{sec:analyticSelftest}

Let's say we are given a self-test candidate formal Bell expression $S$ with the SOS method, such that for any implementation
\begin{equation}
    S = C.1 - \sum_{i\in I} N_i^2.
\end{equation}
If an implementation with the state $\ket{\psi}$ gives $\bra{\psi} \Hat S \ket{\psi} = C$, then for all $i$ we get $\bra{\psi} \Hat N_i^2 \ket{\psi} = 0$ since all of those are positive. Threefore, we obtain: 
\begin{equation}
    \forall \, i \, , \ \Hat N_i \ket{\psi} = 0.
\end{equation}
This is a good starting point to prove a self-test because from an equation on the average value of operators we derived equations that define the action of the operators on the state. This observation has been used in several self-testing proofs~\cite{Supic20}. The idea is to first obtain anti-commutation relations of operators acting on the state.

For instance, considering the singlet example of section II.C, when an implementation gives the maximal value $\bra{\psi} \Hat S(b) \ket{\psi} = C(b)$ we obtain:
\begin{subequations}
    \begin{align}
        & \Hat M_{1}^{(1)} \ket{\psi} = \frac{\Hat M_{1}^{(2)}+\Hat M_{2}^{(2)}}{2\cos(b)} \ket{\psi}, \\
        & \Hat M_{2}^{(1)} \ket{\psi} = \frac{\Hat M_{1}^{(2)}-\Hat M_{2}^{(2)}}{2\sin(b)} \ket{\psi}.
    \end{align}
\end{subequations}
By using the first equation twice and the fact that $\left( \Hat M_{x}^{(1)}\right)^2 = \id$, $\left( \Hat M_{y}^{(2)}\right)^2=\id$ we can obtain $\{\Hat M_{1}^{(2)},\Hat M_{2}^{(2)}\} \ket{\psi} = 2\cos(2b)\ket{\psi}$. Likewise we can obtain $\{\Hat M_{1}^{(1)},\Hat M_{2}^{(1)}\} \ket{\psi}=0$.

These relations can then be further used to derive an extraction map on the measured state and compare the result to the target state~\cite{Supic20}.

\subsection{Self-testing proof for the $\ket{\phi^+}$ inequalities} \label{sec:ProofPhiP}

Let us here prove the self-test result of $\ket{\phi^+}$ for the Bell expression defined in \cref{bell_phi+S}. We suppose that an implementation with state $\ket{\psi}$ and measurements $\Hat M_{x}^{(i)}$ saturates the inequalities \cref{eq:allineqPhiP} for some real values $0< b_1 < a_2 < b_2 < \pi$.

We know from the sum of square decomposition of $S(a_2,b_1,b_2)$ that: 
\begin{equation}
        C(a_2,b_1,b_2)-\Hat{S}(a_2,b_1,b_2)=(\Hat{Z}_A-\Hat{Z}_B)^2 + \frac{1}{f(a_2,b_1,b_2)}\left( \cot(a_2)(\Hat{Z}_A-\Hat{Z}_B) + (\Hat{X}_A-\Hat{X}_B) \right)^2
\end{equation}
holds for any implementation, where $\Hat{Z}_A,\Hat{X}_A,\Hat{Z}_B,\Hat{X}_B$ are given by \cref{eq:SingletPaulis}. Since the statistics we consider saturate the inequality, the average value of each square is $0$ and thus:
\begin{subequations}
    \begin{align}
        (\Hat Z_A-\Hat Z_B)\ket{\psi} &= 0,\\
        (\cot(a_1)(\Hat Z_A-\Hat Z_B) + (\Hat X_A-\Hat X_B))\ket{\psi} &= 0.
    \end{align}
\end{subequations}
From $\Hat Z_A^2=\left(\Hat M_{1}^{(1)}\right)^2=\id$ we can deduce $\Hat Z_B^2\ket{\psi}=\ket{\psi}$ and then using $\left(\Hat M_{1}^{(2)}\right)^2=\left(\Hat M_{2}^{(2)}\right)^2=\id$: 
\begin{equation}
    \{\Hat M_{1}^{(2)},\Hat M_{2}^{(2)}\} \ket{\psi} = 2\cos(b_1-b_2)\ket{\psi}.
\end{equation}
This allows us to get $\Hat X_B^2\ket{\psi}=\ket{\psi}$ and:
\begin{equation}
    \{\Hat Z_B,\Hat X_B\} \ket{\psi} = 0.
\end{equation}
Finally, since we also have by linearity $(\Hat X_A-\Hat X_B)\ket{\psi}=0$, we get $\Hat X_A^2\ket{\psi}=\ket{\psi}$ and $\{\Hat Z_A,\Hat X_A\}\ket{\psi}=0$.

It has been shown before that those (anti-commutation) relations on the control operators $(\Hat Z_A,\Hat X_A,\Hat Z_B,\Hat X_B)$ are enough to self-test the singlet and the action of the measurements (see \cite{McKague12}).

\subsection{Self-testing proof for the family $S_{\theta,b}^n$} \label{sec:ProofPartial}
Suppose that an implementation of a Bell experiment, with state $\ket{\psi}$ and measurements $\Hat M_{x}^{(i)}$ for $x\in\{1,2\}$, $i\in\{1,...,n\}$, gives statistics compatible with $<\Hat S_{\theta,b}^n>_{\ket{\psi}} = 2(1+\lambda^2)$, where $S_{\theta,b}^n$ is defined in \cref{eq:multibellineq}. Due to the sum of squares decomposition given in \cref{eq:sosmoresettings}, we obtain
\begin{subequations}
\begin{align} \label{eq:relation1}
    \left(\Hat M_{1}^{(i)}-\frac{\Hat M_{1}^{(n)}+\Hat M_{2}^{(n)}}{2\cos(b)}\right) \ket{\psi}&= 0,\ \ \forall i < n,\\
\label{eq:relation2}
    \left( \prod_{i=1}^{n-1}\Hat M_{2}^{(i)} \right. - \left.\left[ s_{2\theta}\frac{\Hat M_{1}^{(n)}-\Hat M_{2}^{(n)}}{2\sin(b)} +c_{2\theta}\prod_{i=1}^{n-1}\Hat M_{2}^{(i)}\cdot\frac{\Hat M_{1}^{(n)}+\Hat M_{2}^{(n)}}{2\cos(b)} \right] \right)\ket{\psi} &= 0.
\end{align}
\end{subequations}
Let us define
\begin{align}
        \Hat Z^{(i)}&=\Hat M_{1}^{(i)}, & \Hat X^{(i)}&=\Hat M_{2}^{(i)}, & \Hat P_{\pm}^{(i)}&=\frac{1\pm \Hat Z^{(i)}}{2}, &\ \ \forall i < n\nonumber\\
        Z^{(n)} &= \frac{\Hat M_{1}^{(n)}+\Hat M_{2}^{(n)}}{2\cos(b)}, & \Hat X^{(n)} &= \frac{\Hat M_{1}^{(n)}-\Hat M_{2}^{(n)}}{2\sin(b)},& \Hat P_{\pm}^{(n)}&=\frac{1\pm \Hat Z^{(n)}}{2}.&
\end{align}

For $i<n$, we have $\left(\Hat Z^{(i)}\right)^2 = \id$ from the way we defined them and from \cref{eq:relation1} we have $\Hat Z^{(i)}\ket{\psi}=\Hat Z^{(n)}\ket{\psi}$. Thus we can derive first $\left(\Hat Z^{(n)}\right)^2\ket{\psi}=1$ and:
\begin{equation}
    \begin{split}
        \Hat P_{\pm}^{(i)}\Hat P_{\mp}^{(n)}\ket{\psi}&=0, \\
        \Hat P_{\pm}^{(i)}\Hat P_{\pm}^{(n)}\ket{\psi}&=\Hat P_{\pm}^{(n)}\ket{\psi}
    \end{split} \tag{$*$}
\end{equation}
Moreover, from the defition of $\Hat Z^{(n)}$ and $\Hat X^{(n)}$, we always have $\{\Hat Z^{(n)},\Hat X^{(n)}\}=0$, and thus: 
\begin{equation}
    \Hat X^{(n)}\Hat P_{-}^{(n)}\ket{\psi}=\Hat P_{+}^{(n)}\Hat X^{(n)}\ket{\phi}\tag{$**$}
\end{equation}
Finally, \cref{eq:relation2} allows us to write: 
\begin{equation}
    \Hat X^{(n)}\ket{\psi}=\frac{1}{s_{2\theta}}\prod_{i=1}^{n-1}\Hat X^{(i)}\cdot(1-c_{2\theta}\Hat Z^{(n)})\ket{\psi}\tag{$***$}
\end{equation}

Then we can define the following extraction map, called the swap gate:
\begin{equation}
\ket{0...0}\otimes\ket{\psi}\longrightarrow \sum_{\{a_i\}\in\{0,1\}^n} \prod_{i=1}^{n}\left(\Hat X^{(i)}\right)^{a_i} \prod_{i=1}^{n}\Hat P_{a_i}^{(i)} \ket{a_1...a_{n}}\otimes\ket{\psi}
\end{equation}
Using $(*)$ this map gives: 
\begin{equation}
    \ket{0...0}\otimes \Hat P_{+}^{(n)}\ket{\psi} + \ket{1...1}\otimes \prod_{i=1}^{n-1}\Hat X^{(i)} \cdot \Hat X^{(n)} \Hat P_{-}^{(n)}\ket{\psi}
\end{equation}
Then we can write:
\begin{align}
        \prod_{i=1}^{n-1}\Hat X^{(i)} \cdot \Hat X^{(n)} \Hat P_{-}^{(n)}\ket{\psi} & \underset{\phantom{}(**)}{=} \prod_{i=1}^{n-1}X^{(i)} \cdot \Hat P_{+}X^{(n)}\ket{\psi} \\
        & \underset{[\Hat X^{(i)},\Hat Z^{(n)}]=0}{=} \Hat P_{+}^{(n)} \prod_{i=1}^{n-1}\Hat X^{(i)} \cdot \Hat X^{(n)}\ket{\psi} \\
        & \underset{(***)}{=} \Hat P_{+}^{(n)} \prod_{i=1}^{n-1}X^{(i)} \cdot\frac{1}{s_{2\theta}}\prod_{i=1}^{n-1}\Hat X^{(i)}\cdot(1-c_{2\theta}\Hat Z^{(n)})\ket{\psi}\\
        & \underset{\left(\Hat X^{(i)}\right)^2=\id}{=} \frac{1}{s_{2\theta}} \Hat P_{+}^{(n)}(1-c_{2\theta}\Hat Z^{(n)})\ket{\psi} \\
        & \underset{\left(\Hat Z^{(n)}\right)^2\ket{\psi}=\id}{=} \frac{1-c_{2\theta}}{s_{2\theta}}\Hat P_{+}^{(n)}\ket{\phi} = t_{\theta}\Hat P_{+}^{(n)}\ket{\psi}.
\end{align}
And finally the state is mapped by the swap gate to:
\begin{equation}
(\ket{0...0}+t_\theta\ket{1...1})\otimes \Hat P_{+}^{(n)}\ket{\psi} = \ket{\text{GHZ}_{n,\theta}} \otimes \ket{\textbf{junk}}
\end{equation}

\subsection{Self-testing proof for the family $S_{\theta,b_1,b_2}$} \label{sec:ProofPartial2}

Suppose that an implementation of a Bell experiment in a bipartite scenario with two inputs and two outputs per party, with state $\ket{\psi}$ and measurements $ \Hat M_{x}^{(i)}$ for $x\in\{0,1\}$, $i\in\{1,2\}$, gives statistics compatible with $<\Hat S_{\theta,b_1,b_2}>_{\ket{\psi}} = 2(1+\lambda_1^2+\lambda_2^2)$, where $S_{\theta,b_1,b_2}$ is defined in \cref{eq:ineqmoresettings}. Due to the sum of squares decomposition given in \cref{eq:sosmoresettings}, we obtain
\begin{subequations}
\begin{align} \label{eq:relation11}
    \left(\Hat{Z}_A - \Hat{Z}_B \right) \ket{\psi}&= 0,\\
\label{eq:relation12}
    \left[  \lambda_1 (\Hat{X}_A - s_{2\theta} \Hat{X}_B - c_{2\theta}\Hat{X}_A \Hat{Z}_B) + \lambda_2 (\id - s_{2\theta} \Hat{X}_A \Hat{X}_B - c_{2\theta} \Hat{Z}_B)\right] \ket{\psi}&= 0
\end{align}
\end{subequations}
where $\Hat{Z}_A,\Hat{X}_A,\Hat{Z}_B,\Hat{X}_B$ are given by: 
\begin{subequations}
    \begin{align}
        & \Hat{Z}_A= \Hat M_{1}^{(1)}, \\
        & \Hat{X}_A = \Hat M_{2}^{(1)} \\
        & \Hat{Z}_B=- \frac{\sin(b_2)\Hat M_{1}^{(2)} - \sin(b_1)\Hat M_{2}^{(2)}}{\sin(b_1-b_2)}, \\
        & \Hat{X}_B= \frac{\cos(b_2)\Hat M_{1}^{(2)} - \cos(b_1)\Hat M_{2}^{(2)}}{\sin(b_1-b_2)}.
    \end{align}
\end{subequations}
From their definition, we can derive that $\Hat{Z}_A^2 = \Hat{X}_A^2 = \id$ and thus the equation \cref{eq:relation12} gives:
\begin{equation}
    \Hat{Z}_B^2\ket{\psi}=\ket{\psi}.
\end{equation}
From this we can derive that $s_{b_1}s_{b_2} \{\Hat M_{1}^{(2)},\Hat M_{2}^{(2)}\} \ket{\psi} = (s_{b_1}^2 + s_{b_2}^2 - s_{b_1-b_2}^2)\ket{\psi}$ thus granting the following additional properties:
\begin{subequations}
\begin{align}
    \Hat{X}_B^2\ket{\psi}&=\ket{\psi}, \\
    \{\Hat{Z}_B,\Hat{X}_B\} \ket{\psi}&=0.
\end{align}
\end{subequations}

Let us introduce the following two operators:
\begin{subequations}
    \begin{align}
        \Hat U &= \Hat{X}_A - s_{2\theta} \Hat{X}_B - c_{2\theta}\Hat{X}_A \Hat{Z}_B,\\
        \Hat V &= \id - s_{2\theta} \Hat{X}_A \Hat{X}_B - c_{2\theta} \Hat{Z}_B. 
    \end{align}
\end{subequations}
The second equation \cref{eq:relation12} then becomes
\begin{equation}
    (\lambda_1 \Hat U + \lambda_2 \Hat V) \ket{\psi} = 0
\end{equation}
We can take the square of this equation and use the previous properties on $\Hat Z_B$, $\Hat X_B$ to derive
\begin{equation}
    0 = (\lambda_1 \Hat U + \lambda_2 \Hat V)^2 \ket{\psi} = 2(\lambda_1^2 + \lambda_2 ^2) \Hat V\ket{\psi} + 4\lambda_1\lambda_2 \Hat U \ket{\psi}.
\end{equation}

Now the two previous equations can be linearly combined to obtain:
\begin{equation}
    2(\lambda_1^2 - \lambda_2 ^2) \Hat V \ket{\psi} = 0
\end{equation}
Now we need to compute the term $\lambda_1^2 - \lambda_2 ^2$. This can be done using the fact that $(\lambda_1^2 - \lambda_2 ^2)^2 = (\lambda_1^2 + \lambda_2 ^2)^2 - 4(\lambda_1 \lambda_2)^2$ where both terms are specified in \cref{eq:lambdasexpressions}. We obtain 
\begin{equation}
    (\lambda_1^2 - \lambda_2 ^2)^2 = \frac{4s_{b_1}^2s_{b_2}^2}{(c_{4\theta}-c_{2b_1})(c_{4\theta}-c_{2b_2})}.
\end{equation}
This term is not vanishing for the choice of settings that were made and thus
\begin{equation}
    \Hat V \ket{\psi} = 0.
\end{equation}

Let us now define
\begin{equation}
    \Hat P_{A/B, \pm}^{(i)}=\frac{1\pm \Hat{Z}_{A/B}}{2}\\
\end{equation}
We introduce the swap gate of two qubits:
\begin{equation}
    \ket{00}\otimes\ket{\psi}\longrightarrow \sum_{a,b\in\{0,1\}^n} \Hat{X}_A^a \Hat{X}_B^b \Hat P_{A,a} P_{B,b} \ket{ab}\otimes\ket{\psi}
\end{equation}
The equation \cref{eq:relation11} imposes that $\Hat P_{A,a} \Hat P_{B, b} \ket{\psi} = \delta_{ab} \Hat P_{B,b} \ket{\psi}$ and thus the result of the swap gate is:
\begin{equation}
    \Hat P_{B,0} \ket{00}\otimes\ket{\psi} + \Hat{X}_A \Hat{X}_B \Hat P_{B,1} \ket{11}\otimes\ket{\psi}
\end{equation}
Let us take a closer look at the second member of the above equation. Due to the anti-commutation of $\Hat{Z}_B$ and $\Hat X_B$, and the equation $\Hat V \ket{\psi}= 0$, we obtain:
\begin{subequations}
\begin{align}
    \Hat{X}_A \Hat{X}_B \Hat P_{B,1}\ket{\psi} & =\Hat P_{B,0} \Hat{X}_A \Hat{X}_B \ket{\psi} \\
    & = \Hat P_{B,0} \frac{\id - c_{2\theta} \Hat Z_B}{s_{2\theta}} \ket{\psi} \\
    & = t_\theta \Hat P_{B,0} \ket{\psi}.
\end{align}
\end{subequations}
Finally, the result of the swap gate is given by
\begin{equation}
    (\ket{00}+t_\theta\ket{11})\otimes \Hat P_{B,0} \ket{\psi} = \ket{\phi_\theta} \otimes \ket{\textbf{junk}}.
\end{equation}

\end{document}